\documentclass[11pt]{article}
\usepackage{epsfig}
\usepackage{color}
\usepackage{authblk}
\usepackage{graphicx}
\usepackage{amsmath,amsfonts}
\usepackage[margin=1in]{geometry}
\usepackage{array}
\usepackage{caption}
\usepackage{tikz}
\def\begeq{\begin{equation}}
\def\endeq{\end{equation}}

\def\begeqar{\begin{eqnarray}}
\def\endeqar{\end{eqnarray}}

\def\<{\langle}
\def\>{\rangle}

\newcommand{\be}{\begin{equation}}
\newcommand{\ba}{\begin{eqnarray}}
\newcommand{\ea}{\end{eqnarray}}
\newcommand{\ee}{\end{equation}}

\newcommand{\XXX}[1]{}

\def\begeq{\begin{equation}}
\def\endeq{\end{equation}}

\def\begeqar{\begin{eqnarray}}
\def\endeqar{\end{eqnarray}}

\def\({\left(}		\def\){\right)}
\def\[{\left[}          \def\]{\right]}

\title{The continuum limit of $a_{N-1}^{(2)}$ spin chains}

\author[1]{Eric Vernier\thanks{evernier@sissa.it}}
\author[2,3,4]{Jesper Lykke Jacobsen\thanks{jesper.jacobsen@ens.fr}}
\author[4,5]{Hubert Saleur\thanks{hubert.saleur@cea.fr}}

\affil[1]{SISSA and INFN, Sezione di Trieste, via Bonomea 265, I-34136, Trieste, Italy}
\affil[2]{LPTENS, \'Ecole Normale Sup\'erieure -- PSL Research University, \newline 24 rue Lhomond, F-75231 Paris Cedex 05, France}
\affil[3]{Sorbonne Universit\'es, UPMC Universit\'e Paris 6, CNRS UMR 8549, \newline F-75005 Paris, France} 
\affil[4]{Institut de Physique Th\'eorique, CEA Saclay, F-91191 Gif-sur-Yvette, France}
\affil[5]{USC Physics Department, Los Angeles CA 90089, USA}

\begin{document}

\maketitle

\begin{abstract}
Building on our previous work for $a_2^{(2)}$ and $a_3^{(2)}$ we explore systematically the continuum limit of gapless $a_{N-1}^{(2)}$ vertex models and spin chains. We find the existence of three possible regimes. Regimes I and II  for $a_{2n-1}^{(2)}$ are  related with $a_{2n-1}^{(2)}$ Toda, and described by $n$  compact bosons. Regime I for  $a_{2n}^{(2)}$ is related with $a_{2n}^{(2)}$ Toda and involves $n$ compact bosons, while regime II is related instead with $B^{(1)}(0,n)$ super Toda, and involves in addition  a single Majorana fermion. The most interesting is regime III, where {\sl non-compact} degrees of freedom appear, generalising the emergence of the Euclidean black hole CFT in the $a_{2}^{(2)}$ case. For $a_{2n}^{(2)}$ we find a continuum limit made of $n$ compact and $n$ non-compact bosons, while for $a_{2n-1}^{(2)}$ we find $n$ compact and $n-1$ non-compact bosons. We also find deep relations between  $a_{N-1}^{(2)}$  in regime III and the gauged WZW models  $SO(N)/SO(N-1)$.  

\end{abstract}

\bigskip

\section{Introduction}

The study of the  continuum (or scaling) limits of integrable spin chains is a topic that remains of central importance in theoretical physics, with potential applications  in condensed matter physics and quantum field theory, and more recently in the context of the AdS/CFT correspondence. Identifying these  continuum limits seems a priori a simple technical exercise. The chains are indeed solvable by the Bethe Ansatz, and there is a well-defined procedure, once the ground state and basic excitations are understood, to extract the central charge and critical exponents in an almost rigorous fashion.

The first works in this area quickly proposed, using this strategy, field theories associated, for instance, with integrable chains based on the fundamental representation for all the Lie algebras, including the twisted ones \cite{deVega,deVega1}. Unfortunately, it turned out that the---quite natural---structure of the ground state postulated in these early works was in fact not correct. A more detailed analysis \cite{Nienhuis}, based on a lot of numerics, showed that, even for some of the lowest-rank cases such as $a_2^{(2)}$, various regimes were possible, some of which  exhibiting surprising patterns of roots in their ground states. As fas as we know, no general classification of these patterns has been proposed up to now. Moreover, in several of the regimes, the patterns give rise to considerable technical difficulties, making the numerical or analytical study of the Bethe Ansatz equations very difficult, and hindering a correct identification of the continuum limit.

The $a_2^{(2)}$ case corresponds to the Izergin-Korepin or 19-vertex model \cite{IK81} which is equivalent to a spin-one model of dilute loops on the square lattice \cite{Nienhuis}. Despite its long history, some of its important physical features were only fully understood quite recently \cite{VJS1}. Most notably, the $a_2^{(2)}$ model was shown to exhibit an unexpected `regime III' where the continuum limit is a {\sl non-compact} conformal field theory (CFT) of central charge $c=2$, the so-called Euclidian black hole sigma model \cite{Witten91,Dijkgraaf92} with $SL(2,\mathbb{R})/U(1)$ symmetry.

The emergence of a non-compact CFT, with associated continuous spectrum of critical exponents, was almost unheard of in the field of Bethe Ansatz and quantum spin chains. Quantum spin chains involving finite-dimensional representations of a classical Lie algebra have, in general,  a compact continuum limit, with a discrete set of exponents. But to our knowledge, there is no theorem preventing the emergence of non-compact continuum limits, even if the spins are in finite-dimensional representations, at least  when the `Hamiltonians' are non-Hermitian, which is generally the case in the context of integrable spin chains and $q$-deformations.

The $a_3^{(2)}$ case is related to two Potts models coupled by their energy operator \cite{AuYangPerk,MartinsNienhuis,FJ08}, and allows as well a realisation in terms of loops. Subsequent analysis of the $a_3^{(2)}$ model \cite{VJS2} demonstrated that it exhibits various regimes like the $a_2^{(2)}$ model, including one similar to the `regime III' with a non-compact continuum limit, this time with central charge $c=3$.
A major motivation for the present work was to extend the analysis to the whole $a_{N-1}^{(2)}$ series and to ascertain if non-compact
degrees of freedom are generically present, and if so, how many.

Non-compact CFTs are a subject of high interest in particular for their potential condensed matter applications, which include a variety of geometrical problems, or the description of critical points in $2+1$ dimensional non-interacting disordered electronic systems (such as the IQHE plateau transition: see \cite{BWZ} and references therein) . The possibility of analyzing  these theories using controllable lattice models \cite{IFC11} (as opposed to spin chains involving infinite-dimensional spin representations) is certainly very exciting. Subtle aspects, such as the density of states or the emergence of discrete states in the black hole sigma model \cite{Troost,RibSch}, have already been investigated using lattice techniques \cite{IJS12,VJS1,VJS3}, and there will obviously be much room for progress once other non-compact CFTs have been identified as  low-energy limits of other compact, non-Hermitian spin chains.

This paper is the continuation of two previous works \cite{VJS1,VJS2} on $a_2^{(2)}$ and $a_3^{(2)}$, respectively. The technical difficulties in the analysis of the Bethe Ansatz equations increase very rapidly with the ranks of the algebras, but we will nonetheless provide a general understanding of the continuum limit of $a_{N-1}^{(2)}$ spin chains in their three basic regimes---usually called I, II and III. While regimes I and II are certainly interesting, although they involve rather well-known ingredients, regime III gives rise to a family of non-compact conformal field theories generalising the Euclidian black hole sigma model. While we shall discuss here the main features of these theories, their detailed study will await further work. 

We start out in section~\ref{sec:lattmod} by defining the vertex models of interest in terms of their integrable $\check{R}$-matrix.
We focus on the second solution $\check{R}^{(2)}$ of the Yang-Baxter equations that corresponds to the $a_{N-1}^{(2)}$ models. We
recall their Bethe Ans\"atze and discuss the existence of three regimes. The physics of the regimes I, II and III is established in turn
in the following sections~\ref{sec:I}--\ref{sec:III}. Using an example driven approach---and some numerical assistance---we find in
particular the structure of Bethe roots in the ground state, count the number of compact and non-compact degrees of freedom, and 
identify the (imaginary) Toda theories corresponding to the integrable massive deformations. A chart of our main conclusions can be found in
table~\ref{summ}. We conclude the paper, in section~\ref{sec:conc}, by a summary of our findings and an outlay of directions for
further work. A discussion of a free-field representation of the $SO(N)_k/SO(N-1)_k$ cosets is relegated to appendix~\ref{sec:app}.

\section{Integrable lattice models based on $a_{N-1}^{(2)}$}
\label{sec:lattmod}

\subsection{$\check{R}$-matrices}

Integrable vertex models and spin chains based on the twisted affine Lie algebras $a_{N-1}^{(2)}$ have appeared sporadically in the literature, motivated largely by the technical difficulties associated with the twisting. These models---in the fundamental representation case, to which we restrict now---are, however, also  interesting for applications, in particular because they provide 
\cite{GZ,GM} a second family of solutions of the Yang-Baxter equation with  (quantum deformation of) $so(N)$ symmetry. This observation generalises the simple fact that there are {\sl two} solutions of the Yang-Baxter equation for the three-dimensional `spin-one' representation of $U_qsl(2)$: the Fateev-Zamolodchikov model \cite{FZ80} and the Izergin-Korepin model \cite{IK81}.

The technical point is that one can Baxterise in two different ways the  Birman-Murakami-Wenzl (BMW) algebra \cite{BirmanWenzl,Murakami} associated with $so(N)$.  Since this point will be crucial later in our analysis of the regime III of these models, we discuss it further. 

The first Baxterisation is the one associated with $so(N)^{(1)}$, and we write the corresponding $\check{R}$-matrix as $\check{R}^{(1)}$. It is given by the first formula found in \cite{McKay}%
\footnote{After a correction in eq.\ (2.4): $\check{P}\rightarrow P$.}
\begin{equation}
\check{R}^{(1)} \propto \check{P}_S-{q^2x-1\over q^2-x}\check{P}_A+{q^2x-1\over q^2-x}{q^{N-2}x-1\over q^{N-2}-x}\check{P}_0 \,,
\end{equation}
where we have defined 
\begin{eqnarray}
\check{P}_i&=&{\cal P} P_i \,, \nonumber\\
\check{P}_S&=&P_S \,, \nonumber\\
\check{P}_A&=&-P_A \,, \nonumber\\
\check{P}_0&=&P_0 \,,
\end{eqnarray}
obeying $P_S+P_A+P_0=I$, with $I$ being the identity operator. Here ${\cal P}$ is the permutation operator, whereas $P_i$ with $i=S,A,0$ denote the orthogonal projectors onto the symmetric, antisymmetric and trivial representation, respectively. As usual, $q$ is the quantum group deformation parameter, and $x$ denotes the spectral parameter.

The braid limit is $x\to 0$, leading to 
\begin{equation}
\left. \check{R}^{(1)} \right|_{x \to 0} \propto \check{P}_S+q^{-2}\check{P}_A+q^{-N}\check{P}_0 \,.
\end{equation}
We define the braid generators
\begin{eqnarray}
B&=&q P_S-q^{-1}P_A+q^{1-N}P_0 \,, \nonumber\\
B^{-1}&=&q^{-1}P_S-qP_A+q^{N-1}P_0 \,. \label{Bope}
\end{eqnarray}
They satisfy the Kauffman skein relation  
\begin{equation}
B-B^{-1}=(q-q^{-1})(1-E) \,, \label{skein}
\end{equation}
where we have introduced the braid monoid
\begin{equation}
E=\left(1+[N-1]\right)P_0
\end{equation}
and the $q$-deformed (quantum) numbers
\begin{equation}
 [n] = \frac{q^{n}-q^{-n}}{q-q^{-1}} \,.
\end{equation}

In addition to (\ref{skein}), the defining relations of the $so(N)$  BMW algebra are the braid relations
\begin{eqnarray}
 B_i B_{i+1} B_i &=& B_{i+1}B_iB_{i+1} \,, \nonumber\\
 B_i B_j &=& B_j B_i \quad \mbox{for } |i-j| \ge 2 \,; \label{braidrel}
\end{eqnarray}
the idempotent relation
\begin{equation}
 E_i^2 = \left(1+[N-1]\right)E_i \,;
\end{equation}
the delooping relations
\begin{eqnarray} 
 B_i E_i = E_i B_i &=& q^{1-N} E_i \,, \nonumber\\
 E_i B_{i \pm 1} E_i &=& q^{N-1} E_i \,;
\end{eqnarray}
and finally the tangle relations
\begin{eqnarray}
 E_i E_{i \pm 1} E_i &=& E_i \,, \nonumber\\
 B_i B_{i \pm 1} E_i &=& E_{i \pm 1} E_i \,. \label{tanglerel}
\end{eqnarray}
%
All these relations can be depicted diagramatically, using the well-known representations of $E_i$ and $B_i$ in terms of contractions and over-passings of adjacent strands.

It is straightforward to rewrite the $\check{R}$-matrix as
\begin{eqnarray}
\check{R}^{(1)}&\propto& (q^{-1}-q)x\left(x-q^{N-2}\right) I
+(x-1)\left(x-q^{N-2}\right)B+\left(q-q^{-1}\right)x(x-1)E \,, \nonumber\\
\check{R}^{(1)}&\propto& I +{x-1\over x+1}{q^{N-2}+x\over q^{N-2}-x}E+{1-x\over 1+x}{1\over q-q^{-1}}\left(B+B^{-1}\right) \,, \label{R1def}
\end{eqnarray}
where of course the proportionality coefficients are irrelevant. 

Now, there is another solution of the Yang-Baxter equations with the same symmetry, the same underlying BMW algebra, and acting in the same product of fundamental representations. This second $R$-matrix  reads
\begin{equation}
\check{R}^{(2)}=\check{P}_S-{q^2x-1\over q^2-x}\check{P}_A+{q^Nx+1\over q^N+x}\check{P}_0 \,,
\end{equation}
with the same braid limit as before
\begin{equation}
\left. \check{R}^{(2)} \right|_{x \to 0} =\check{P}_S+q^{-2}\check{P}_A+q^{-N}\check{P}_0 \,.
\end{equation}
It leads to expressions similar to (\ref{R1def}):
\begin{eqnarray}
\check{R}^{(2)}&\propto& (q^{-1}-q)x\left(x+q^{N}\right) I
+(x-1)\left(x+q^{N}\right)B+\left(q-q^{-1}\right)x(x-1)E \,, \nonumber\\
\check{R}^{(2)}&\propto & I +{x-1\over x+1}{q^{N}-x\over q^{N}+x}E+{1-x\over 1+x}{1\over q-q^{-1}}\left(B+B^{-1}\right) \,.
\label{newsol}
\end{eqnarray}
In the modern classification of solutions of the Yang-Baxter equation, this second solution is associated with $a_{N-1}^{(2)}$. This $\check{R}^{(2)}$-matrix coincides with that of $a_{N-1}^{(2)}$ given in \cite{Bazhanov,Jimbo}. For a detailed study of $\check{R}$-matrices based on twisted quantum affine algebras, see \cite{Delius/Gould/Zhang}. 

In the remainder of this paper the parity of $N$ will play an important role---as is generally the case for  CFTs and integrable models with $so(N)$ symmetry.
When $N=2n+1$ is odd, the $\check{R}^{(2)}$-matrix is $U_q(b_n)$ invariant \cite{Kuniba,Artz1,Artz2}. 
The situation for $N=2n$ is more complicated. The $\check{R}^{(2)}$-matrix as it was described here---being obtained from the $so(N)$ BMW algebra---must clearly be $U_q(d_n)$ invariant. On the other hand, $U_q(c_n)$ invariance is claimed in part of the literature \cite{Artz1,Artz2}. Moreover, the Bethe Ansatz for the associated vertex model is usually indexed with the eigenvalues of the $c_n$ Cartan generators \cite{Reshetikhin}. We will follow this convention here.%
\footnote{We note that in \cite{Kuniba} a different $a_{2n-1}^{(2)}$ $\check{R}$-matrix has been proposed, which has $U_q(c_n)$ symmetry. There is a strong suspicion \cite{Kunibaprivate} that this $\check{R}$-matrix and the one in \cite{Bazhanov,Jimbo} lead to identical Bethe equations in the periodic case.}
We have checked explicitly for small sizes and various ranks that the usual Bethe equations for $a_{2n-1}^{(2)}$ \cite{Reshetikhin} do indeed give the correct levels for the model based on the $\check{R}^{(2)}$-matrix.%
\footnote{This Bethe Ansatz was rederived `from first principles' in \cite{GM} in the more general case of $sl(n|m)^{(2)}$.}

Lattice models of clear physical interest are well-known for $a_2^{(2)}$, which is related in particular with a spin-one $O(n)$ loop model%
\footnote{The parameter $n$ in this notation is related with the $q$-deformation, and has nothing to do with the rank of an algebra.}
on the square lattice \cite{Nienhuis}, which is based on the Izergin-Korepin vertex model. The same spin chain---albeit in a different regime \cite{VJS2016}---is related to the chromatic polynomial on the triangular lattice \cite{Baxter86} and from there to several geometrical models of the Potts and $O(n)$ loop-model types \cite{VJS2016}.
More recently, a physical interpretation of the $a_3^{(2)}$ model in terms of a two-colour loop model was provided \cite{AuYangPerk,MartinsNienhuis,FJ08}. There is so far no such interpretation, to our knowledge, for higher values of $N$. 

\subsection{The Bethe Ansatz} 

The Bethe equations are well-known  (see for instance \cite{GM1} and references therein). They are of rank $n$ for both $a_{2n-1}^{(2)}$ and $a_{2n}^{(2)}$ and read (using the parameterisation $q=e^{i\gamma}$, and for periodic boundary conditions):
\begin{enumerate}
 \item For $a_{2n-1}^{(2)}$: with $j=2,3,\ldots,n-2$ 
 \begin{eqnarray}
\left({\sinh (\lambda^1-i{\gamma\over 2})\over \sinh(\lambda^1+i{\gamma\over 2})}\right)^L&=&\prod_{\lambda^{1'}}^{m_1} {\sinh(\lambda^1-\lambda^{1'}-i\gamma)\over\sinh(\lambda^1-\lambda^{1'}+i\gamma)}\prod_{\lambda^{2}}^{m_2} {\sinh(\lambda^1-\lambda^2+i{\gamma\over 2})\over\sinh(\lambda^1-\lambda^2-i{\gamma\over 2})} \,, \nonumber\\
\prod_{\lambda^{j-1}}^{m_{j-1}} {\sinh(\lambda^j-\lambda^{j-1}-i{\gamma\over 2})\over\sinh(\lambda^j-\lambda^{j-1}+i{\gamma\over 2})}&=&\prod_{\lambda^{j'}}^{m_j}{\sinh(\lambda^j-\lambda^{j'}-i\gamma)\over\sinh(\lambda^j-\lambda^{j'}+i\gamma)}\prod_{\lambda^{j+1}}^{m_{j+1}}{\sinh (\lambda^j-\lambda^{j+1}+i{\gamma\over 2})\over\sinh (\lambda^j-\lambda^{j+1}-i{\gamma\over 2})} \,, \nonumber\\
\prod_{\lambda^{n-2}}^{m_{n-2}} {\sinh(\lambda^{n-1}-\lambda^{n-2}-i{\gamma\over 2})\over\sinh(\lambda^{n-1}-\lambda^{n-2}+i{\gamma\over 2})}&=&\prod_{\lambda^{n-1'}}^{m_{n-1}}{\sinh(\lambda^{n-1}-\lambda^{n-1'}-i\gamma)\over\sinh(\lambda^{n-1}-\lambda^{n-1'}+i\gamma)}\prod_{\lambda^{n}}^{m_n}{\sinh 2(\lambda^{n-1}-\lambda^n+i{\gamma\over 2})\over\sinh 2(\lambda^{n-1}-\lambda^n-i{\gamma\over 2})} \,,\nonumber\\
\prod_{\lambda^{n-1}}^{m_{n-1}} {\sinh2 (\lambda^n-\lambda^{n-1}-i{\gamma\over 2})\over\sinh 2(\lambda^n-\lambda^{n-1}+i{\gamma\over 2})}&=&\prod_{\lambda^{n'}}^{m_n} {\sinh 2(\lambda^n-\lambda^{n'}-i\gamma)\over \sinh 2(\lambda^n-\lambda^{n'}+i\gamma)} \,. \label{BetheEqs1}
\end{eqnarray}
\item For $a_{2n}^{(2)}$: with $j=2,3,\ldots,n-1$ 
\begin{eqnarray}
\left({\sinh (\lambda^1-i{\gamma\over 2})\over \sinh(\lambda_1+i{\gamma\over 2})}\right)^L&=&\prod_{\lambda^{1'}}^{m_1} {\sinh(\lambda^1-\lambda^{1'}-i\gamma)\over\sinh(\lambda^1-\lambda^{1'}+i\gamma)}\prod_{\lambda^{2}}^{m_2} {\sinh(\lambda^1-\lambda^2+i{\gamma\over 2})\over\sinh(\lambda^1-\lambda^2-i{\gamma\over 2})} \,, \nonumber\\
\prod_{\lambda^{j-1}}^{m_{j-1}} {\sinh(\lambda^j-\lambda^{j-1}-i{\gamma\over 2})\over\sinh(\lambda^j-\lambda^{j-1}+i{\gamma\over 2})}&=&\prod_{\lambda^{j'}}^{m_j}{\sinh(\lambda^j-\lambda^{j'}-i\gamma)\over\sinh(\lambda^j-\lambda^{j'}+i\gamma)}\prod_{\lambda^{j+1}}^{m_{j+1}}{\sinh (\lambda^j-\lambda^{j+1}+i{\gamma\over 2})\over\sinh (\lambda^j-\lambda^{j+1}-i{\gamma\over 2})} \,, \nonumber\\
\prod_{\lambda^{n-1}}^{m_{n-1}} {\sinh (\lambda^n-\lambda^{n-1}-i{\gamma\over 2})\over\sinh (\lambda^n-\lambda^{n-1}+i{\gamma\over 2})}&=&\prod_{\lambda^{n'}}^{m_n} {\sinh (\lambda^n-\lambda^{n'}-i\gamma)\over \sinh (\lambda^n-\lambda^{n'}+i\gamma)}{\cosh (\lambda^n-\lambda^{n'}+i{\gamma\over 2})\over \cosh (\lambda^n-\lambda^{n'}+i{\gamma\over 2})} \,. \label{BetheEqs2}
\end{eqnarray}
\end{enumerate}
These equations can be obtained from the $a_{2n-1}^{(1)}$ and $a_{2n}^{(1)}$ Bethe equations respectively by a `folding' of the roots \cite{Reshetikhin,deVega}. In both sets of equations, $m_j$ denotes the number of Bethe roots $\lambda^j$ of type $j=1,2,\ldots,n$.
Note that $\lambda^j$ is generally defined modulo $i \pi$, except for $\lambda^n$ in the $a_{2n-1}^{(2)}$ case which is modulo $\frac{i \pi}{2}$ only.

Solving the Bethe equations gives access to the full spectrum (assuming that the Bethe Ansatz is complete) of the general vertex model based on the $\check{R}^{(2)}$ matrix discussed in the foregoing section. Eigenvalues of the transfer matrix are then used to extract the central charge and the critical exponents, via the usual finite-size scaling formulae. It will be convenient in what follows to refer to the anisotropic limit of the vertex model where the logarithmic derivative of the transfer matrix becomes a local Hamiltonian. 
The energy eigenvalues then take the form
\begin{equation}
E=-{\cal N}\sum_{i=1}^{m_1} {\sin\gamma\over \cosh 2\lambda^1_j-\cos\gamma} \,, \label{energy}
\end{equation}
where ${\cal N}$ is a constant depending on normalisation of the Hamiltonian.  Different regimes will correspond to different choices of the {\sl sign} of ${\cal N}$, as well as the value of $\gamma$. For a given sign, the {\sl absolute value} of ${\cal N}$ is then chosen to ensure a relativistic continuum limit (that is, a dispersion relation $e=p$ for low-energy excitations). 

Like in all problems of this sort, it is crucial to perform numerical studies of the lattice model in order to understand which kind of Bethe roots are associated with the ground state and low-energy excitations. The periodic row-to-row transfer matrix has the structure%
\footnote{We henceforth omit the superscript on $\check{R}^{(2)}$.}
\be 
T_L(x)= \mathrm{tr}_a \left( \check{R}_{a,L}(x) \ldots \check{R}_{a,2}(x) \check{R}_{a,1}(x)  \right) \,,
\ee
where each of the $L$ quantum (vertical) spaces as well as the auxilliary (horizontal) space carry the $N$-dimensional fundamental representation of $so(N)$, and $\check{R}$ is given in terms of the algebra generators by (\ref{newsol}). Its explicit form is \cite{Jimbo}
\begin{eqnarray}
\check{R}_{ab}(x) &=& (x-\xi)(x-q^2) \sum_{\substack{\alpha=1 \\ \alpha \neq \alpha'}}^{N} 
\hat{e}^{(a)}_{\alpha \alpha} \otimes \hat{e}^{(b)}_{\alpha \alpha}
+q(x-1)(x-\xi)\sum_{\substack{\alpha,\beta=1 \\ \alpha \neq \beta,\alpha \neq \beta'}}^{N}
\hat{e}^{(a)}_{\beta \alpha} \otimes \hat{e}^{(b)}_{\alpha \beta}  \nonumber \\
&+& x(1-q^2)(x-\xi) \sum_{\substack{\alpha,\beta=1 \\ \alpha < \beta,\alpha \neq \beta'}}^{N} \hat{e}^{(a)}_{\alpha \alpha} \otimes \hat{e}^{(b)}_{\beta \beta}
+(1-q^2)(x-\xi)  \sum_{\substack{\alpha,\beta=1 \\ \alpha > \beta,\alpha \neq \beta'}}^{N} \hat{e}^{(a)}_{\alpha \alpha} \otimes \hat{e}^{(b)}_{\beta \beta} \nonumber \\
&+& \sum_{\alpha ,\beta =1}^{N} d_{\alpha, \beta} (x)
\hat{e}^{(a)}_{\alpha' \beta} \otimes \hat{e}^{(b)}_{\alpha \beta'} \,,
\label{Rexplicit}
\end{eqnarray}
where the notation $\alpha' \equiv N+1-\alpha$ is used, and $\hat{e}^{(a)}_{\alpha \beta}$ (resp.\ $\hat{e}^{(b)}_{\alpha \beta}$) denotes the matrix acting non-trivially on the tensorand labelled $a$ (resp.\ $b$),
such that $\left(\hat{e}_{\alpha \beta}\right)_{\mu \nu} = \delta_{\alpha \mu} \delta_{\beta \nu}$. Moreover $\xi = -q^N$, whilst $d_{\alpha\beta}(x)$ has the form
\begin{equation}
d_{\alpha, \beta} (x) = \left \lbrace
\begin{array}{ll}
 q(x -1)(x -\xi) +x(q^2 -1)(\xi -1) &
 \mbox{for } \alpha=\beta=\beta' \,, \\
 (x -1)\left[ (x -\xi)q^{2} +x(q^2 -1) \right] &
 \mbox{for } \alpha=\beta \neq \beta' \,, \\
 (q^{2 }-1)\left[ \xi(x -1) q^{\bar{\alpha}-\bar{\beta}} -\delta_{\alpha ,\beta'} (x -\xi) \right] &
 \mbox{for } \alpha < \beta \,, \\
 (q^{2 }-1) x \left[ (x -1) q^{\bar{\alpha}-\bar{\beta}} -\delta_{\alpha ,\beta'} (x -\xi) \right] &
 \mbox{for } \alpha > \beta \,, \\
 \end{array} \right.
\end{equation}
where 
\be
\bar{\alpha} 
= \begin{cases}  \alpha+\frac{1}{2} \quad &\text{for}~ 1\leq \alpha< \frac{N+1}{2} \,, \\
\alpha \quad &\text{for}~  \alpha =\frac{N+1}{2} \,, \\
\alpha-\frac{1}{2} \quad &\text{for}~ \frac{N+1}{2}< \alpha \leq N \,. \\ \end{cases}
\ee

From the quantum integrability of the model, the transfer matrices for different values of the spectral parameter $x$ commute  and therefore share the same set of eigenvectors. This does not prevent level crossings, and the set of states determining the largest transfer matrix eigenvalues may vary with $x$. More precisely, for each $N$ there are two `isotropic' values $x_{\pm} = \mathrm{e}^{2 i \left(\frac{N \gamma}{4} \mp \frac{\pi}{4} \right)}$, corresponding to local maxima of the transfer matrix eigenvalues, and which are described by a different physics in the sense that they are not dominated by the same eigenstates. From the Hamiltonian point of view, these correspond to opposite signs in the definition of the energy $E$, namely $x_\pm$ correspond to respectively ${\cal N} >0$ and ${\cal N}<0$ in (\ref{energy}). As already announced above this gives rise to different regimes, whose precise description we give below (section \ref{ref:regimes}).

For $a_{2n}^{(2)}$ we have the value of the Cartan generators in the $b_n \equiv so(2n+1)$ subalgebra 
\begin{eqnarray}
h_1 &=& L-m_1 \,, \nonumber\\
h_j &=& m_{j-1}-m_j \quad \mbox{for } j=2,3,\ldots,n \,.
\end{eqnarray}
For $a_{2n-1}^{(2)}$ we have similarly the Cartan generators in the $c_n\equiv sp(2n)$ subalgebra:
\begin{eqnarray}
 h_1 &=& L-m_1\nonumber\\
 h_j &=& m_{j-1}-m_j \quad \mbox{for } j=2,3,\ldots,n-1 \,, \nonumber\\
 h_{n} &=& m_{n-1}-2m_n \,,
\end{eqnarray}
where the $m_j$ are the numbers of Bethe roots in  (\ref{BetheEqs1}) and (\ref{BetheEqs2}).  We have restricted to the case of $L$ even  to avoid parity and spurious twist effects. For all cases studied explicitly, we checked that  the ground state lies in the singlet sector with all the $h_j=0$.

\subsection{The regimes}
\label{ref:regimes}

For $a_{2n-1}^{(2)}$, the transformation $\gamma\to \pi-\gamma$ combined with a shift of roots $\lambda^1$ by ${i\pi\over 2}$ is equivalent to changing the sign of the coupling constant:  ${\cal N}\to -{\cal N}$ in (\ref{energy}). It is therefore enough to study the region $\gamma\in [0,{\pi\over 2}]$ for both signs of ${\cal N}$.  We will see that this gives rise to three regimes, but two have essentially identical physical properties:
$$
\begin{tabular}{>{$}l<{$}  >{$}l<{$}  >{$}l<{$}}
 \gamma \in \left[ 0,{\pi\over 2} \right] \,, &  {\cal N}<0 \,: & \text{regime I} \\[1mm]
 \gamma \in \left[ {\pi\over 2n},{\pi\over 2} \right] \,,  & {\cal N}>0 \,: & \text{regime I'} \\[1mm]
 \gamma \in \left[ 0,{\pi\over 2n} \right] \,, & {\cal N}>0 \,: & \text{regime III} \\
\end{tabular}
$$
For $a_{2n}^{(2)}$, there is no such symmetry, since  the last  ($\cosh$) term in the Bethe equations (\ref{BetheEqs2}) involves ${\gamma\over 2}$. Accordingly, there are in fact three totally different regimes:
$$
\begin{tabular}{>{$}l<{$}  >{$}l<{$}  >{$}l<{$}}
 \gamma \in [0,\pi] \,, & {\cal N}<0 \,: & \text{regime I} \\[1mm]
 \gamma \in \left[ {\pi\over 2n+1}, \pi \right] \,, & {\cal N}>0 \,: & \text{regime II} \\[2mm] 
 \gamma \in \left[0,{\pi\over 2n+1} \right] \,, & {\cal N}>0 \,: & \text{regime III} \\
\end{tabular}
$$

Throughout the remainder of the paper, we will use the denomination regime II to refer also to regime I' of $a_{2n-1}^{(2)}$, checking a posteriori that in the latter case it is nothing but the analytic continuation of regime I. 

We now turn to a detailed analysis of all three regimes. 

\section{Regime I}
\label{sec:I}

\subsection{The case $a_{2n}^{(2)}$ in regime I}

This corresponds to ${\cal N}<0$ and $\gamma\in [0,\pi]$. The Bethe roots in the ground state organise themselves into the pattern $\lambda^1=x^1+{i\pi\over 2},\lambda^2=x^2,\lambda^3=x^3+{i\pi\over 2},\lambda^4=x^4\ldots$, where the $x^j$ are real.
In all that follows, we will define  Fourier transforms  via
\begin{equation}
f(\omega)=\int {{\rm d}\lambda\over 2\pi} \, e^{i \lambda \omega} f(\lambda),
\end{equation}
and use the basic formulae
\begin{eqnarray}
{{\rm d}\over {\rm d}\lambda}\ln {\sinh(\lambda+i\alpha)\over \sinh(\lambda-i\alpha}&=&\int_{-\infty}^\infty {\rm d}\omega \, \cos\omega\lambda
{\sinh \omega \left({\pi\over 2}-\alpha\right)\over \sinh{\omega\pi\over 2}} \,, \nonumber\\
{{\rm d}\over {\rm d}\lambda}\ln {\cosh(\lambda-i\alpha)\over \cosh(\lambda+i\alpha)}&=&\int_{-\infty}^\infty {\rm d}\omega \, \cos\omega\lambda
{\sinh \omega \alpha\over \sinh{\omega\pi\over 2}} \,.
\end{eqnarray}
The Bethe equations in the thermodynamic  limit, restricting to the types of roots that appear in the ground state,%
\footnote{There are, as usual, more types of roots, but these do not play an essential role in the understanding of the continuum limit.}
have then the  simple form (with $j=2,3,\ldots,n-1$)
\begin{eqnarray}
\rho_1+\rho_1^h&=&{\sinh \omega\gamma/2\over \sinh \omega\pi/2}+{\sinh\omega({\pi\over 2}-\gamma)\over \sinh \omega\pi/2}\rho_1+{\sinh \omega\gamma/2\over \sinh \omega\pi/2}\rho_2\nonumber\\
\rho_j+\rho_j^h&=&{\sinh \omega\gamma/2\over \sinh \omega\pi/2}\rho_{j-1}+
{\sinh\omega({\pi\over 2}-\gamma)\over \sinh \omega\pi/2}\rho_j+
{\sinh \omega\gamma/2\over \sinh \omega\pi/2}\rho_{j+1}\nonumber\\
\rho_n+\rho_n^h&=&{\sinh \omega\gamma/2\over \sinh \omega\pi/2}\rho_{n-1}+\left[
{\sinh\omega({\pi\over 2}-\gamma)\over \sinh \omega\pi/2}+{\sinh\omega\gamma/2\over\sinh \omega\pi/2}\right]\rho_n\label{a2ncont}
\end{eqnarray}
where $\rho_j$ and $\rho_j^h$ are densities of Bethe roots and holes per unit length for the $j$'th type of excitations. There are $n$ massless modes, and the central charge is $c=n$. 

We can rewrite this in the compact, symbolic form
\begin{equation}
\pmb{\rho}+\pmb{\rho}^h=\pmb{s}+\pmb{K}\star\pmb{\rho} \,, \label{symbol}
\end{equation}
where the densities $\pmb{\rho}$, $\pmb{\rho}^h$ and the source term $\pmb{s}$ are column vectors, and the
interaction kernel $\pmb{K}$ is a matrix. Taking $\pmb{K}$ at zero frequency produces
\begin{equation}
1-\pmb{K}(0)={\gamma\over\pi}\left(\begin{array}{rrrrrr}
 2 & -1& 0 & \cdots & 0 & 0 \\
-1 & 2 &-1 & \cdots & 0 & 0 \\
 \vdots & \ddots & \ddots & \ddots &  & \vdots\\[-1mm]
 \vdots & & \ddots & \ddots & \ddots & \vdots \\
 0 & 0 & \cdots & -1 & 2 &-1 \\
 0 & 0 & \cdots & 0 & -1 & 1
 \end{array}\right)\equiv \pmb{R} \,. \label{Kzerofreq}
\end{equation}
Here $\pmb{R}$ is equal to ${\gamma\over\pi}$ times the symmetrised Cartan matrix  ($\pmb{\alpha}_i \cdot \pmb{\alpha}_j$) of the $b_n$ algebra, where $\pmb{ \alpha}_1,\ldots,\pmb{ \alpha}_n$ are the roots of $b_n$; notice that this holds even for $n=1$. Conformal weights corresponding to excitations made out of holes (in numbers $\delta m_i$) and global shifts of the Fermi seas (with $\delta d_i$ roots `backscattered from left to right') are given by \cite{Suzuki} 
\begin{equation}
\Delta+\bar{\Delta}={1\over 4}  \delta \pmb m \cdot \pmb R \cdot \delta \pmb m+ \delta \pmb d \cdot \pmb R^{-1} \cdot \delta \pmb d\label{genfor}
\end{equation}
The continuum limit is therefore a set of $n$ compact bosons $\phi_i$. The exact compactification rules deserve further study, since $b_n$ is not simply laced, but we will not pursue this matter here---except to stress that in this regime, there are no indications of further fermionic degrees of freedom. Observe that the conformal weights associated with pure hole excitations read
\begin{equation}
\Delta+\bar{\Delta}
={\gamma\over 4\pi}\left[(\delta m_1)^2+(\delta m_2-\delta m_1)^2+\ldots+(\delta m_{n-1}-\delta m_n)^2\right]\label{confweight2n}
\end{equation}
and can be naturally associated with  vertex operators $V\equiv \exp\left(\sum_{i=1}^n \delta m_i \pmb{\alpha}_i \cdot \pmb{\phi}\right)$. 

It is well-known \cite{deVega,deVega1,ResSal} that an integrable spin chain provides not only a lattice discretisation of a  conformal field theory, but also the discretisation of an integrable massive deformation thereof. The latter is obtained  by {\sl staggering the bare spectral parameter}, so the source terms in the Bethe equations (\ref{BetheEqs2}) are modified:
\begin{equation}
\left( {\sinh\left(\lambda^1_j-i{\gamma\over 2}\right)\over  
\sinh\left(\lambda^1_j+i{\gamma\over 2}\right)} \right)^L\to \left( {\sinh\left(\lambda^1_j-\Lambda-i{\gamma\over 2}\right)\over  
\sinh\left(\lambda^1_j-\Lambda+i{\gamma\over 2}\right)} \right)^{L/2}\left( {\sinh\left(\lambda^1_j+\Lambda-i{\gamma\over 2}\right)\over  
\sinh\left(\lambda^1_j+\Lambda+i{\gamma\over 2}\right)} \right)^{L/2}
\end{equation}
($\Lambda$ is a real parameter) with a similar staggering in the transfer matrix/time evolution \cite{ResSal}. The field theoretic limit is obtained close to vanishing energy/momentum. This requires taking $\Lambda$ large, and focusing on  a region where the source term for the density of holes is dominated by the poles nearest the origin: we will discuss this in more detail below for some examples.  Masses and scattering matrices can then be determined, and the massive field theory identified.

The result for the $a_{2n}^{(2)}$ model in regime I  is that 
staggering  produces the imaginary  $a_{2n}^{(2)}$ Toda theory (for general discussion of Toda theories, see \cite{BCDS})  with the action
\begin{eqnarray}
S&=&\int {1\over 2} (\partial_\mu\pmb{\phi} \cdot \partial_\mu\pmb{\phi})+ g\left[e^{-2i\beta\phi_1}+2\sum_{i=1}^{n-1}e^{i\beta(\phi_i-\phi_{i+1})}+2e^{i\beta\phi_{n}}\right]\nonumber\\
&=&\int {1\over 2} (\partial_\mu\pmb{\phi} \cdot \partial_\mu\pmb{\phi})+ g\left(e^{-2i\beta\pmb{\alpha}_0 \cdot \pmb{\phi}}+2\sum_{i=1}^{n}e^{i\beta\pmb{\alpha}_i \cdot \pmb{\phi}_i}\right) \,, \nonumber\\
\label{Todageni}\end{eqnarray}
where $\pmb{\alpha}_0$ satisfies 
\begin{equation}
\pmb{\alpha}_0+2(\pmb{\alpha}_1+\pmb{\alpha}_2+\ldots+\pmb{\alpha}_{n})=0 \,.
\end{equation}
This Toda theory is based on the   $a_{2n}^{(2)}$ affine root system (the $\pmb{e}_{i}$ being as usual a set of orthonormal vectors)
\begin{eqnarray}
\pmb{\alpha}_{i} &=& \pmb{e}_{i}-\pmb{e}_{i+1} \quad \mbox{for } i=1,2,\ldots,n-1 \,, \nonumber \\
\pmb{\alpha}_n &=& \pmb{e}_n \,, \nonumber \\
\pmb{\alpha}_{0} &=& -2\pmb{e}_1\label{rootsbn} \,.
\end{eqnarray}
The corresponding Dynkin diagram is shown in figure~\ref{Dynkin1}.
The $\pmb K$-matrix (\ref{Kzerofreq}) and the form of the conformal weights (\ref{confweight2n}) are compatible with the exponentials in (\ref{Todageni}), provided that
\begin{equation}
{\beta^2\over 8\pi}={\gamma\over 2\pi} \,. \label{dimcond}
\end{equation}

 \begin{figure}[htbp]
\begin{center}
\begin{tikzpicture}
\node at (-0.4,0) {$\alpha_0$};
\draw (0,0.08) -- (1,0.08);
\draw (0,-0.08) -- (1,-0.08);
\draw (0.9,0.15) -- (1.08,0) -- (0.9,-0.15);
\node at (1.5,0) {$\alpha_{1}$};
\draw (1.9,0) -- (2.5,0);
\draw[dashed] (2.5,0) -- (4,0);
\draw (4,0) -- (4.7,0);
\node at (5.2,0) {$\alpha_{n-1}$};
\draw (5.7,0.08) -- (6.7,0.08);
\draw (5.7,-0.08) -- (6.7,-0.08);
\draw (6.6,0.15) -- (6.78,0) -- (6.6,-0.15);
\node at (7.2,0.) {$\alpha_{n}$};
\end{tikzpicture}
\caption{Dynkin diagram of $a_{2n}^{(2)}$.}
\label{Dynkin1}
\end{center}
\end{figure}

\subsection{The case $a_{2n-1}^{(2)}$ in regime I}

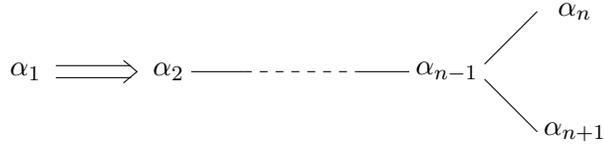
\begin{figure}[htbp]
\begin{center}
\begin{tikzpicture}
\node at (-0.4,0) {$\alpha_1$};
\draw (0,0.08) -- (1,0.08);
\draw (0,-0.08) -- (1,-0.08);
\draw (0.9,0.15) -- (1.08,0) -- (0.9,-0.15);
\node at (1.5,0) {$\alpha_2$};
\draw (1.8,0) -- (2.5,0);
\draw[dashed] (2.5,0) -- (4,0);
\draw (4,0) -- (4.7,0);
\node at (5.2,0) {$\alpha_{n-1}$};
\draw (5.7,0.1) -- (6.4,0.8);
\node at (6.9,0.8) {$\alpha_{n}$};
\draw (5.7,-0.1) -- (6.4,-0.8);
\node at (6.9,-0.8) {$\alpha_{n+1}$}; 
\end{tikzpicture} 
\caption{Dynkin diagram of $a_{2n-1}^{(2)}$.}
\label{Dynkin}
\end{center}
\end{figure}

Regime I is observed for ${\cal N}<0$ and $\gamma\in [0,{\pi\over 2}]$. 
The Bethe roots for $a_{2n-1}^{(2)}$ exhibit a pattern of alternation between imaginary parts $0$ and ${\pi\over 2}$:  $\lambda^1=x^1+{i\pi\over 2},\lambda^2=x^2,\lambda^3=x^3+{i\pi\over 2},\lambda^4=x^4,\ldots$, except for the last roots which have imaginary part ${\pi\over 4}$: $\lambda^n=x^n+{i\pi\over 4}$.  The Bethe equations in the thermodynamic limit read (with $j=2,3,\ldots,n-2$)
\begin{eqnarray}
\rho_1+\rho_1^h&=&{\sinh {\omega\gamma \over 2}\over \sinh {\omega\pi \over 2}}+{\sinh\omega({\pi\over 2}-\gamma)\over \sinh {\omega\pi \over 2}}\rho_1+{\sinh {\omega\gamma \over 2} \over \sinh {\omega\pi \over 2}}\rho_2 \,, \nonumber\\
\rho_j+\rho_j^h&=&{\sinh {\omega\gamma \over 2} \over \sinh {\omega\pi \over 2}}\rho_{j-1}+
{\sinh\omega({\pi\over 2}-\gamma)\over \sinh {\omega\pi \over 2}}\rho_j+
{\sinh {\omega\gamma \over 2} \over \sinh {\omega\pi \over 2}}\rho_{j+1} \,, \nonumber\\
\rho_{n-1}+\rho_{n-1}^h&=&{\sinh {\omega\gamma \over 2} \over \sinh {\omega\pi \over 2}}\rho_{n-2}+
{\sinh\omega({\pi \over 2}-\gamma)\over \sinh {\omega\pi \over 2}}\rho_{n-1}+{\sinh {\omega\gamma \over 2} \over \sinh {\omega\pi \over 4}}\rho_{n-1} \,, \nonumber\\
\rho_{n}+\rho_{n}^h&=&{\sinh {\omega\gamma \over 2} \over \sinh {\omega\pi \over 4}}\rho_{n-1}+
{\sinh\omega({\pi \over 4} -\gamma)\over \sinh {\omega\pi \over 4}}\rho_{n} \,. \label{a2n-1cont}
\end{eqnarray}
The $\pmb{K}$-matrix obeys 
\begin{equation}
1-\pmb{K}(0)={\gamma\over\pi}\left(\begin{array}{rrrrrr}
 2 & -1& 0 & \cdots & 0 & 0 \\
-1 & 2 &-1 & \cdots & 0 & 0 \\
 \vdots & \ddots & \ddots & \ddots &  & \vdots\\[-1mm]
 \vdots & & \ddots & \ddots & \ddots & \vdots \\
 0 & 0 & \cdots & -1 & 2 &-2 \\
 0 & 0 & \cdots & 0 & -2 & 4
\end{array}\right)\equiv \pmb{R} \,,
\end{equation}
and $\pmb R$ coincides now with ${\gamma\over\pi}$ times  the symmetrised Cartan matrix  ($\pmb{\alpha}_i \cdot \pmb{\alpha}_j$) of the $c_n$ algebra. The central charge is $c=n$ as for $a_{2n}^{(2)}$, and equation (\ref{genfor}) applies as well. The conformal weights associated with pure hole excitations read
\begin{equation}
\Delta+\bar{\Delta}={\gamma\over 4\pi}\left[(\delta m_1)^2+(\delta m_2-\delta m_1)^2+\ldots+(\delta m_{n-2}-\delta m_{n-1})^2+(\delta m_{n-1}-2\delta m_{n})^2\right]
\end{equation}
and can be naturally associated with  vertex operators $V\equiv \exp\left(\sum_{i=1}^n \delta m_i \pmb{\alpha}_i \cdot \pmb{\phi}\right)$. The staggering  produces the imaginary 
$a_{2n-1}^{(2)}$ Toda  theory with action 
\begin{equation}
S = \int {1\over 2} (\partial_\mu\pmb{\phi} \cdot \partial_\mu\pmb{\phi})+
g\left[e^{i\beta(\phi_{1}-\phi_2)}+e^{-i\beta(\phi_{1}+\phi_2)}+2\sum_{j=2}^{n-1}
e^{i\beta(\phi_j-\phi_{j+1})}+e^{2i\beta\phi_n}\right]\label{Todagenii}
\end{equation}
based on  the $a_{2n-1}^{(2)}$ affine root system given by
\begin{eqnarray}
 \pmb{\alpha}_{i} &=& \pmb{ e}_{i}-\pmb{ e}_{i+1} \quad \mbox{for } i=1,2,\ldots,n-1 \,, \nonumber \\
 \pmb{ \alpha}_n &=& 2\pmb{ e}_n \,, \nonumber \\
 \pmb{ \alpha}_{0} &=& -\pmb{ e}_{1}-\pmb{ e}_2
\end{eqnarray}
obeying
\begin{equation}
\pmb{ \alpha}_0+\pmb{\alpha}_1+2(\pmb{ \alpha}_2+\ldots+\pmb{\alpha}_{n-1})+\pmb{ \alpha}_n=0 \,.
\end{equation}
The corresponding Dynkin diagram is shown in figure \ref{Dynkin}. The $n$ roots ${\pmb \alpha}_1,\ldots,{\pmb \alpha}_n$ are those of the algebra $c_n$. The correspondence requires the same condition (\ref{dimcond}) as before.

We now turn to a series of examples to justify our claims.

\subsection{Example 1: $a_2^{(2)}$}
\label{sec:a22RI}

This example has a long history \cite{IK81,Nienhuis}, and was discussed in great detail in the appendix of our first paper \cite{VJS1}. We recall its main features here for completeness.

The Bethe equations (\ref{a2ncont}) read now simply
\begin{equation}
\rho+\rho^h={\sinh {\omega\gamma\over 2}\over\sinh{\omega\pi\over 2}}+{2\sinh{\omega\over 4}(\pi-\gamma)\cosh{
\omega\over 4}(\pi-3\gamma)\over \sinh{\omega\pi\over 2}}\rho \,.
\end{equation}
The `physical equations' obtained by putting the density of excitations over the physical ground state on the right  are then
\begin{equation}
\rho+\rho^h={\cosh{\omega\over 4}(\pi-\gamma)\over \cosh{3\omega\over 4}(\pi-\gamma)}-{\sinh{\omega\over 2}(\pi-\gamma)\cosh{\omega\over 2}(\pi-3\gamma)\over\sinh{\omega\gamma\over 2}\cosh {3\omega\over 4}(\pi-\gamma)}\rho^h \,. \label{BethephysI}
\end{equation}
After staggering, these equations inherit the new source term in Fourier space
\begin{equation}
{\cos\Lambda\omega \cosh{\omega\over 4}(\pi-\gamma)\over \cosh{3\omega\over 4}(\pi-\gamma)}  \,.
\end{equation}
When going back to real space, this becomes a complicated expression in terms of the rapidity $\lambda$ of the holes. The field theoretic limit is obtained close to vanishing energy/momentum. This requires taking $\Lambda$ large, and focusing on  a region where the source term is dominated by the poles nearest the origin, here $\omega=\pm {2i\over 3}{\pi\over (\pi-\gamma)}$. In this limit, the source term  is proportional to $\exp\left[-{2\Lambda\over 3}{\pi\over \pi-\gamma}\right]\cosh {2\over 3}{\pi\over \pi-\gamma}\lambda$. This leads to the mass scale
\begin{equation}
M\propto \exp\left[-\Lambda{2\over 3-{3\gamma\over\pi}}\right] \,, \label{massscale}
\end{equation}
and the physical rapidity is $\theta={2\over 3}{\pi\over \pi-\gamma}\lambda$.
 
The low-energy limit of the staggered model then  corresponds to an integrable relativistic quantum field theory. The latter is easily identified, once one recognises  the kernel in the Bethe equation (\ref{BethephysI}) as the (logarithmic derivative of the) $S$-matrix \cite{Takacs,Birgit1} for the Bullough-Dodd model \cite{DB77,ZS79} with (non-real) action
\begin{equation}
S=\int {1\over 2} (\partial_\mu\phi)^2+g(e^{-2i\beta\phi}+e^{i\beta\phi}) \,, \label{BDact}
\end{equation}
where one should set 
\begin{equation}
{\beta^2\over 8\pi}={\gamma\over 2\pi} \,.
\end{equation}
In our units, this is the conformal weight of $e^{i\beta\phi}$. Knowing the  action in the continuum limit allows us to obtain the relationship between the bare coupling $g$ in (\ref{BDact})  and the staggering $e^{-\Lambda}$ on the lattice. Imagine indeed computing perturbatively the ground state energy of the model with action (\ref{BDact}). This will expand in powers of $g^3$ since only three-point functions involving one insertion of the first exponential and two insertions of the  second one will contribute. By dimensional analysis, it follows that $[g]=[\hbox{length}]^{-2+{\beta^2\over 8\pi}}=[\hbox{length}]^{-{2(1-{\gamma\over\pi})}}$.  Comparing with (\ref{massscale}), we get thus that
\begin{equation}
g\propto e^{-{4\over 3}\Lambda} \,. \label{couprel}
\end{equation}
From $[g]=[\hbox{length}]^{-{2(1-{\gamma\over\pi})}}$ we see that the coupling becomes dimensionless for $\gamma=\pi$, in agreement with the natural boundary of regime I.

\subsection{Example 2: $a_3^{(2)}$ }
\label{sec:a32RI}

This case also has a fairly long history \cite{AuYangPerk,MartinsNienhuis,FJ08} and was treated in some detail in our second paper \cite{VJS2}.

The ground state does not involve complexes, and is given by configurations of the type
\begin{equation}
\lambda^1=x^1+{i\pi\over 2} \,, \qquad \lambda^2=x^2+{i\pi\over 4} \,.
\end{equation}
We recall that $\lambda^1$ is defined modulo $i\pi$ and $\lambda^2$ is defined modulo ${i\pi\over 2}$.
The equations for the real parts read
\begin{eqnarray}
\left({\cosh (x^1-{i\gamma\over 2})\over \cosh(x^1+{i\gamma\over 2})}\right)^L&=&\prod_{x^{1'}} {\sinh (x^1-x^{1'}-i\gamma)\over \sinh(x^1-x^{1'}+i\gamma)}\prod_{x^2} {\cosh 2(x^1-x^2+{i\gamma\over 2})\over \cosh 2(x^1-x^2-{i\gamma\over 2})} \,, \nonumber\\
\prod_{x^1} {\cosh 2(x^2-x^1-{i\gamma\over 2})\over \cosh 2(x^2-x^1+{i\gamma\over 2})}&=&\prod_{x^{2'}} {\sinh 2(x^2-x^{2'}-i\gamma)\over \sinh 2(x^2-x^{2'}+i\gamma)} \,.
\end{eqnarray}
Denoting by $\rho_1$ and $\rho_2$ the corresponding densities, we recover  the equations for densities in the thermodynamic limit (\ref{a2n-1cont}) for $n=2$:
\begin{eqnarray}
\rho_1+\rho_1^h&=&{\sinh {\omega\gamma\over 2}\over \sinh {\omega\pi\over 2}}+{\sinh \omega({\pi\over 2}-\gamma)\over \sinh {\omega\pi\over 2}}\rho_1+{\sinh {\omega\gamma\over 2}\over \sinh {\omega\pi\over 4}}\rho_2 \,, \nonumber\\
\rho_2+\rho_2^h&=&{\sinh {\omega\gamma\over 2}\over \sinh {\omega\pi\over 4}}\rho_1+{\sinh \omega({\pi\over 4}-\gamma)\over \sinh {\omega\pi\over 4}}\rho_2 \,.
\end{eqnarray}
Writing the Bethe equations symbolically in the usual form (\ref{symbol}),
we have the following $\pmb{K}$-matrix at zero frequency
\begin{equation}
\pmb{K}(0)=\left(\begin{array}{cc}
1-{2\gamma\over\pi} &{2\gamma\over\pi}\\
{2\gamma\over\pi}&1-{4\gamma\over\pi}\end{array}\right) \,,
\end{equation}
and thus
\begin{equation}
1-\pmb{K}(0)={\gamma\over\pi}\left(\begin{array}{cc} 2&-2\\-2&4\end{array}\right) \,,
\end{equation}
which is equal to ${\gamma\over\pi}$ times the symmetrised Cartan matrix of $c_2=b_2$.
This means we expect the low-energy spectrum to have the contribution coming from holes 
\begin{equation}
\Delta+\bar{\Delta}={\gamma\over 4\pi} \left[(\delta m_1)^2+(\delta m_1-2\delta m_2)^2\right] \,.
\label{eq:a32Igaps}
\end{equation}
It may now be useful to recast things in terms of the `two-colour' interpretation of the $a_3^{(2)}$ model \cite{FJ08,VJS2}. The fundamental representation of $so(4)$ can be decomposed in terms of $su(2) \times su(2)$, and a basis for the Cartan generators is then given in terms of the longitudinal component of two $su(2)$ spins, $S_z$ and $S_z'$, defined in the basis of equation (\ref{Rexplicit}) as $S_z = \mathrm{diag}(-\frac{1}{2},-\frac{1}{2},\frac{1}{2},\frac{1}{2})$ and $S_z' = \mathrm{diag}(-\frac{1}{2},\frac{1}{2},-\frac{1}{2},\frac{1}{2})$. The correspondence with the number of roots $m_1$ and $m_2$ is given by
\begin{eqnarray}
\delta m_1&=&\hbox{change in the number of $\lambda^1$ roots}=S_z+S_z' \,, \nonumber\\
\delta m_2&=&\hbox{change  in the number of $\lambda^2$ roots}=S_z' \,,
\end{eqnarray}
and the gaps (\ref{eq:a32Igaps}) take the form
\begin{equation}
\Delta+\bar{\Delta}={\gamma\over 2\pi} (S_z^2+S_z'^2) \,.
\end{equation} 
%

The physical equations are now
\begin{eqnarray}
\rho_1+\rho_1^h&=&{\cosh \omega({\pi\over 4}-{\gamma\over 2})\over \cosh \omega({3\pi\over 4}-\gamma)}-{\sinh {\omega\pi\over 2}\over 2\sinh {\omega\gamma\over 2}\cosh \omega({3\pi\over 4}-\gamma)}\rho_2^h
-{\sinh {\omega\over 2}(\pi-\gamma)\cosh \omega(\gamma-{\pi\over 4})\over 
\sinh {\omega\gamma\over 2}\cosh \omega({3\pi\over 4}-\gamma)}\rho_1^h \,, \nonumber\\
\rho_2+\rho_2^h&=&{1\over 2\cosh \omega({3\pi\over 4}-\gamma)}-{\sinh {\omega\pi\over 2}\over 2\sinh {\omega\gamma\over 2}\cosh \omega({3\pi\over 4}-\gamma)}\rho_1^h-{\sinh \omega({\pi\over 4}-{\gamma\over 2})\cosh \omega({\pi\over 2}-\gamma)\over \sinh {\omega\gamma\over 2}\cosh \omega({3\pi\over 4}-\gamma)}\rho_2^h \,. \nonumber 
\\
\label{physa32}
\end{eqnarray}
Staggering the bare spectral parameter leads to  a massive integrable QFT  which can be identified%
\footnote{A similar example of rank two is discussed in \cite{BWKSsu3} for the $a_2$ spin chains.}
with the imaginary $d_3^{(2)}$ Toda theory \cite{GandenbergerMcKay}.
Indeed, in the latter reference we find the following data (we use the subscripts GK from the authors' initials to refer to these). First, the masses of the solitons are
\begin{eqnarray}
M_a &\propto& \sin {a\pi\over 3}\left({1\over 2}-{1\over 3\lambda_{\rm GK}}\right) \qquad \mbox{with } a=1,2 \,; \\
\lambda_{\rm GK} &\equiv& {4\pi\over \beta_{\rm GK}^2}-{4\over 3} \,.
\end{eqnarray}
We also introduce, following the same reference \cite{GandenbergerMcKay}, $\omega_{\rm GK}={2\pi\over \beta_{\rm GK}^2}-1$. We now take the $S_{11}$ soliton-soliton scattering matrix element given in their eq.~(18) and rewrite it in terms of Fourier integrals. This gives 
\begin{equation}
\ln F_{11}=\int_{-\infty}^\infty {{\rm d}t\over t} \, {\sinh \mu_{\rm GK} t\over \sinh {t\over 2}\cosh {t\over 2}(3\omega_{\rm GK}+1)}\sinh {t\omega_{\rm GK}\over 2}\cosh t\omega_{\rm GK} \,,
\end{equation}
where 
\begin{equation}
\mu_{\rm GK}=-{3i\lambda_{\rm GK}\theta\over 2\pi}
\end{equation}
($\theta$ being the rapidity), and 
\begin{equation}
 \omega_{\rm GK}={2\pi\over \beta_{\rm GK}^2}-1 \,.
\end{equation}
To compare with our results,  we observe that the Fourier transform of the source term in our equations (\ref{physa32}) is (we still call $\lambda$ the generic real parts of the roots in what follows)
\begin{equation}
 \int_{-\infty}^\infty {\rm d}\omega \, e^{i\lambda \omega} {1\over \cosh \omega({3\pi\over 4}-\gamma)}={\pi\over {3\pi\over 4}-\gamma}{1\over\cosh \left({{\pi\over 2}\over {3\pi\over 4}-\gamma}\lambda\right)} \,.
\end{equation}
Massless excitations will occur at large rapidities, where the source term is thus proportional to $e^{\pm i\theta}$, with $\theta$ denoting the renormalised rapidity:
\begin{equation}
 \theta= {2\pi\over 3\pi-4\gamma}\lambda \,.
\end{equation}
This behaviour, which occurs entirely because of the pole at $i{\pi\over 2}$, leads us immediately to the ratio of the two soliton masses in our model:
 \begin{equation}
 {M_2\over M_1}=2\cos {\pi\over 2}{\pi-2\gamma\over 3\pi-4\gamma} \,.
 \end{equation}
Note however that the soliton with mass $M_1$ in \cite{GandenbergerMcKay} corresponds to holes $\rho_2^h$, and the soliton of mass $M_2$ corresponds to holes $\rho_1^h$, so there is an inversion of labels. Setting 
\begin{equation}
 {M_2\over M_1}=2\cos {\pi\over 3}\left({1\over 2}-{1\over 3\lambda_{GK}}\right) 
\end{equation}
leads to the key identification
\begin{eqnarray}
 \gamma &=& {\beta_{\rm GK}^2\over 4} \,, \nonumber \\
 \lambda_{\rm GK} &=& {4\pi\over \beta_{\rm GK}^2}-{4\over 3}={\pi\over \gamma}-{4\over 3} \,.
\end{eqnarray}
Setting now
\begin{equation}
t\equiv \omega\gamma
\end{equation}
it follows from these mappings that 
\begin{equation}
|\mu_{GK}|t={3\lambda_{GK}\over 2\pi}\theta t= \omega\lambda \,,
\end{equation}
and thus that we can rewrite 
\begin{eqnarray}
\sinh{t\over 2}&=&\sinh {\omega\gamma\over 2} \,, \nonumber\\
\cosh {t\over 2}(3\omega_{GK}+1)&=&\cosh \omega\left({3\pi\over 4}-\gamma\right) \,, \nonumber\\
\sinh {t\omega_{GK}\over 2}&=&\sinh \omega\left({\pi\over 4}-{\gamma\over 2}\right) \,, \nonumber\\
\cosh t\omega_{GK}&=&\cosh \omega\left({\pi\over 2}-\gamma\right) \,,
\end{eqnarray}
so the $\rho_2^h$--$\rho_2^h$ scattering is correctly described (recall the inversion of labels) by $F_{11}$. Similar calculations show the same holds for $F_{12},F_{22}$ as given in \cite{GandenbergerMcKay}. We thus recognise here the massless limit of the $d_n^{(2)}$ Toda theory in the particular case of $d_3^{(2)}=a_3^{(2)}$.

%
%
%
%

To finish the identification, we observe that the pole nearest the real axis provides the following correspondence between the mass scale and the staggering parameter
\begin{equation}
M\propto \exp\left[-\Lambda{2\over 3-{4\gamma\over\pi}}\right] \,.
\end{equation}
On the other hand, if the perturbation  for $a_3^{(2)}$ Toda reads
\begin{equation}
g\left[e^{i\beta(\phi_1-\phi_2)}+e^{-i\beta(\phi_1+\phi_2)}+e^{2i\beta\phi_2}\right] \,,
\end{equation}
we will need, using the same kind of  argument as for $a_2^{(2)}$,
\begin{equation}
[g]\propto [\hbox{length}]^{{2\beta^2\over 3\pi}-2} \,.
\end{equation}
This leads to the following relationship between $g$ and the staggering parameter in the  $a_3^{(2)}$ model:
\begin{equation}
g\propto e^{-{4\over 3}\Lambda}
\end{equation}
(this is in fact  the same relationship as in the $a_2^{(2)}$ case), together with
\begin{equation}
{\beta^2\over 8\pi}
={\gamma\over2\pi} \,.
\end{equation}
%
%
%

Note that, in terms of $\gamma$, the dimension of the bare coupling is obtained via
\begin{equation}
[g]\propto [\hbox{length}]^{{8\over 3}{\gamma\over\pi}-2} \,.
\end{equation}
It becomes dimensionless  when $\gamma={3\pi\over 4}$, suggesting that regime I should have a continuation past ${\pi\over 2}$, as we shall  see below. 

\bigskip
The two foregoing examples fully illustrate the general pattern, with results summarised at the beginning of this section. We have carried out explicitly the analysis of the next two cases, in particular to ascertain the nature of the roots and the spectrum of excitations. We content ourselves by mentioning just a few relevant features below.

\subsection{Example 3: $a_4^{(2)}$}
\label{sec:a42RI}

The ground state is obtained with $\lambda^1=x^1+{i\pi\over 2}$ and $\lambda^2=x^2$, with the continuum equations
\begin{eqnarray}
\rho_1+\rho_1^h&=&{\sinh {\omega\gamma \over 2} \over \sinh {\omega\pi \over 2}}+{\sinh\omega({\pi\over 2}-\gamma)\over \sinh {\omega\pi \over 2}}\rho_1+{\sinh {\omega\gamma \over 2} \over \sinh {\omega\pi \over 2}}\rho_2 \,, \nonumber\\
\rho_2+\rho_2^h&=&{\sinh {\omega\gamma \over 2} \over \sinh {\omega\pi \over 2}}\rho_1+\left[
{\sinh\omega({\pi\over 2}-\gamma)\over \sinh {\omega\pi \over 2}}+{\sinh {\omega\gamma \over 2} \over\sinh {\omega\pi \over 2}}\right]\rho_2 \,,
\end{eqnarray}
so we have
\begin{equation}
 1-\pmb{K}(0)={\gamma\over\pi}\left(\begin{array}{rr}
 2&-1\\
-1&1 \end{array}\right) \,.
\end{equation}
This is equal to ${\gamma\over\pi}$ times the symmetrised Cartan matrix of $b_2$ (of course, the root systems of $b_2$ and $c_2$ are isomorphic, but the distinction between the two algebras is relevant for the higher-rank cases), and the hole part of the finite-size spectrum is given by 
\begin{equation}
\Delta+\bar{\Delta}={\gamma\over 4\pi} \left[(\delta m_1)^2+(\delta m_2-\delta m_1)^2\right] \,.
\end{equation}
The physical equations are of the form
\begin{eqnarray}
\rho_1+\rho_1^h&=&{1\over\cosh \omega{5(\pi-\gamma)\over 4}}+\cdots \,, \nonumber\\
\rho_2+\rho_2^h&=&{\cosh \omega{\pi-\gamma\over 4}\over \cosh \omega{3(\pi-\gamma)\over 4} \cosh \omega{5(\pi-\gamma)\over 4}}+\cdots \,.
\end{eqnarray}
Introducing the usual staggering,  we see  that we will get a scattering theory with two types of solitons  and that the ratio of their masses is independent of $\gamma$, in contrast with the $a_3^{(2)}$ case:
\begin{equation}
{M_2\over M_1}={\sin {2\pi\over 5}\over \sin{\pi\over 5}}={\sin{2\pi\over H}\over \sin{\pi\over H}} \,,
\end{equation}
with $H$ the Coxeter number ($H=2n+1$ for  $a_{2n}^{(2)}$). The mass scale is fixed by the relation at the pole 
\begin{equation}
M\propto \exp\left[-\Lambda {2\over 5-5{\gamma\over\pi}}\right] \,.
\end{equation}
A more detailed analysis of the scattering kernels shows that the equations are describing the 
$a_4^{(2)}$ Toda theory, with perturbation
\begin{equation}
g\left[e^{-2\beta\phi_1}+2e^{i\beta(\phi_1-\phi_2)}+2e^{\beta\phi_2}\right] \,,
\end{equation}
with the---by now usual---result ${\beta^2\over 8\pi}={\gamma\over 2\pi}$, and the relation between the staggering and the coupling is 
\begin{equation}
g\propto e^{-{4\over 5}\Lambda}\propto [\hbox{length}]^{{2\gamma\over\pi}-2} \,.
\end{equation}
It becomes dimensionless for  $\gamma=\pi$. 

Note that in the identification of the scattering theory with the results of \cite{GandenbergerMcKay} the soliton with mass $M_2$ corresponds to holes $\rho_2^h$ and the soliton of mass $M_1$ to holes $\rho_1^h$: in the case of $a_{2n}^{(2)}$ there is no label inversion, in contrast with the case of $a_{2n-1}^{(2)}$ (see section~\ref{sec:a32RI}).

\subsection{Example 4: $a_5^{(2)}$}
\label{sec:a52RI}

The ground state is of the form 
\begin{equation}
\lambda^1=x^1+{i\pi\over 2}, \quad \lambda^2=x^2, \quad \lambda^3=x^3+{i\pi\over 4}
\end{equation}
with the bare Bethe equations
\begin{eqnarray}
\rho_1+\rho_1^h &=& {\sinh {\omega\gamma \over 2} \over \sinh {\omega\pi \over 2}}+{\sinh {\omega\gamma \over 2} \over \sinh {\omega\pi \over 2}} \rho_2+{\sinh \omega({\pi \over 2}-\gamma)\over \sinh {\omega\pi \over 2}} \rho_1 \,, \nonumber\\
\rho_2+\rho_2^h &=& {\sinh {\omega\gamma \over 2} \over \sinh {\omega\pi \over 2}} \rho_1+{\sinh {\omega\gamma \over 2} \over \sinh {\omega\pi \over 4}} \rho_3+{\sinh \omega({\pi \over 2}-\gamma)\over \sinh {\omega\pi \over 2}}\rho_2 \,, \nonumber\\
\rho_3+\rho_3^h &=& {\sinh {\omega\gamma \over 2} \over \sinh {\omega\pi \over 4}} \rho_2+{\sinh \omega({\pi \over 4}-\gamma) \over \sinh {\omega\pi \over 4}} \rho_3 \,.
\end{eqnarray}
One has 
\begin{equation}
1-\pmb{K}(0)={\gamma\over\pi}\left(\begin{array}{rrr}
2&-1&0\\
-1&2&-2\\
0&-2&4
\end{array}\right) \,,
\end{equation}
which is proportional to the symmetrised Cartan matrix of $c_3$. The hole part of the finite-size spectrum thus has the form 
\begin{equation}
\Delta+\bar{\Delta}={\gamma\over\pi}\left[(n_1^2+(n_2-n_1)^2+(n_2-2n_3)^2\right] \,.
\end{equation}
The physical equations have the form
\begin{eqnarray}
\rho_1+\rho_1^h&=&{\cosh\omega(3\pi/4-\gamma)\over \cosh\omega(5\pi/4-3\gamma/2)}+\cdots \,, \nonumber\\
\rho_2+\rho_2^h&=&{\cosh\omega(\pi/4-\gamma/2)\over \cosh\omega(5\pi/4-3\gamma/2)}+\cdots \,, \nonumber\\
\rho_3+\rho_3^h&=&{1\over 2\cosh\omega(5\pi/4-3\gamma/2)}+\cdots \,.
\end{eqnarray}
It follows that the usual staggering now leads to a mass scale, from the nearest pole of the cosh in the denominator, 
\begin{equation}
M\propto \exp\left[-\Lambda {2\over 5-6{\gamma\over \pi}}\right] \,.
\end{equation}
The masses of the three types of solitons, {\sl after inversion of the labels}, are given in this case by
\begin{eqnarray}
{M_2\over M_1} &=& 2\sin \pi{\pi-\gamma\over 5\pi-6\gamma} \,, \nonumber \\
{M_3\over M_1} &=&
2\sin 2 \pi{\pi-\gamma\over 5\pi-6\gamma} \,.
\end{eqnarray}

A more detailed study  suggests that the continuum limit is the   $a_5^{(2)}$ Toda theory with perturbation
\begin{equation}
g\left[e^{-2\beta\phi_1}+2e^{\beta(\phi_1-\phi_{2})}+e^{\beta(\phi_{2}-\phi_3)}+e^{\beta(\phi_{2}+\phi_3)}\right]
\end{equation}
and ${\beta^2\over 8\pi}={\gamma\over 2\pi}$.  Dimensional analysis gives 
\begin{equation}
g\propto e^{-4\Lambda\over 5}\propto [\hbox{length}]^{{12\over 5}{\gamma\over\pi}-2} \,.
\end{equation}
Like for $a_3^{(2)}$ the coupling only becomes dimensionless at $\gamma={5\pi\over 6}$, suggesting the existence of a continuation of the regime. 

\subsection{Remarks}

The imaginary $a_{2n}^{(2)}$ Toda theories have been discussed in \cite{BigRef2}. The $S$-matrices found in that reference can be matched  in detail against the lattice model results, generalising the analysis for $a_2^{(2)}$. One can for instance easily check that the mass ratios are independent of the coupling, as we did for $a_4^{(2)}$. This is related with the theory being self-dual.

Meanwhile, we are not aware of any systematic study of the  $S$-matrices  for imaginary $a_{2n-1}^{(2)}$, except, as discussed in section~\ref{sec:a32RI} above,  for $a_3^{(2)}=d_3^{(2)}$.  The lattice models provide a natural route to obtain these matrices: it is clear from the lattice Bethe Ansatz that the mass ratios will, in general, be coupling dependent (unlike what happens for $a_{2n}^{(2)}$). The  $S$-matrix for the real version of this theory was determined in \cite{DGZ}: some features of this $S$-matrix can be  extrapolated to the complex regime, with results in agreement with the lattice analysis. A similar discussion will be presented in the following section.

Note that in all cases we can write the relationship between the mass scale and the staggering parameter  $\Lambda$  as
\begin{eqnarray}
M&\propto& \exp\left[-\Lambda {2\over H(1-{\gamma\over\pi})}\right] \,, \qquad \quad \mbox{for } a_{2n}^{(2)} \,; \nonumber\\
M&\propto& \exp\left[-\Lambda {2\over H-(H+1){\gamma\over\pi}}\right] \,, \quad \! \mbox{for } a_{2n+1}^{(2)} \,,
\end{eqnarray}
where the Coxeter number  is $H=2n+1$ for both of $a_{2n}^{(2)}$ and $a_{2n+1}^{(2)}$\footnote{We use $a_{2n+!}^{(2)}$ instead of $a_{2n-1}^{(2)}$ in this paragraph so as to have a single Coxeter  for both types of algebras.}.
The correspondence between the Toda coupling and the staggering parameter is then
\begin{eqnarray}
g&\propto& e^{-{4\over H}\Lambda}\propto [\hbox{length}]^{2{\gamma\over\pi}-2} \,, \qquad \quad \mbox{for } a_{2n}^{(2)} \,; \nonumber\\
g&\propto& e^{-{4\over H}\Lambda}\propto [\hbox{length}]^{2{H+1\over H}{\gamma\over\pi}-2} \,, \quad \ \mbox{for } a_{2n+1}^{(2)} \,.
\end{eqnarray}
 The second equation suggests the extension of the regime up to ${\gamma\over\pi}={H\over H+1}={2n+1\over 2n+2}$ for $a_{2n+1}^{(2)}$. 

In the case $a_{2n}^{(2)}$ we have $n$ solitons with masses 
\begin{equation}
{M_a\over M_1}={\sin a{\pi\over H}\over \sin{\pi\over H}} \,, \quad \mbox{for } a=1,2,\ldots,n \,,
\end{equation}
while in the case of $a_{2n+1}^{(2)}$  we found $n+1$ solitons with $\gamma$-dependent masses
\begin{equation}
{M_a\over M_1}=2\sin \left[a\pi{\pi-\gamma\over H\pi-(H+1)\gamma}\right] \,, \quad \mbox{for } a=2,3,\ldots,n+1 \,.
\end{equation}

\section{Regime II}
\label{sec:II}

 \subsection{The case of $a_{2n}^{(2)}$ in regime II}
\label{sec:ane2RII}

This corresponds to ${\cal N}>0$ and  $\gamma\in[{\pi\over 2n+1},\pi]$.
From explicit study of the cases $a_{2}^{(2)}$ and $a_{4}^{(2)}$, we conjecture that the ground state in this regime is described by the following patterns of roots, 
\begin{equation}
\lambda^{\alpha} \simeq  x \pm i \left(\frac{\pi}{4} - \frac{(2(n-\alpha)+1) \gamma }{4}\right)\,, \qquad \mbox{for } \alpha = 1,2,\ldots,n \,,
\end{equation}
where in the first line the notation $\simeq$ means that the real parts $x$ of the different types of roots are only equal up to corrections decreasing exponentially fast with $L$; similarly the imaginary parts are only equal to their asymptotic values up to such corrections.%
\footnote{In fact this roots configuration is strictly valid only for $\gamma$ `not too far' from $\frac{\pi}{2n+1}$: when $\gamma$  increases, some of the imaginary parts go to zero, leading to a merging of the corresponding 2-strings. Typically the corresponding roots then become real, leading to a different set of equations. An explicit example will be treated in the case of $a_4^{(2)}$ (see section \ref{sec:a42RII}), showing that this does not modify the thermodynamic and conformal properties of the continuum limit. The case $a_2^{(2)}$ was discussed previously in \cite{VJS1}.}
In the $L \to \infty$ limit, as these corrections vanish, one can write the first set of Bethe Ansatz equations for $\lambda^1 = x + i \left( \frac{\pi}{4} - \frac{2n-1}{4}\gamma \right)$ (resp. $\lambda^1 = x - i \left( \frac{\pi}{4} - \frac{2n-1}{4}\gamma \right)$) as
\begin{eqnarray}
\left(  
\frac{\sinh \left(x + i \left( \frac{\pi}{4}-\frac{2n+1\gamma}{4}\right)\right)}{\sinh \left(x + i \left(\frac{\pi}{4}+\frac{(2n-3)\gamma}{4}\right)\right)}
\right)^L
\!\!\!\!\! &=& \!\!\!
\prod_{x'} 
\frac{\sinh \left(x-x'\right)}{\sinh \left(x-x' + i \gamma\right)} 
\frac{\sinh \left(x-x'  + i \left(\frac{\pi}{2}-\frac{2n+1}{2}\gamma\right)\right)}{\sinh \left(x-x'  + i \left(\frac{\pi}{2}-\frac{2n-1}{2}\gamma\right)\right)} 
\,, \nonumber \\
\left(  
\frac{\sinh \left(x -i \left(\frac{\pi}{4}+\frac{(2n-3)\gamma}{4}\right)\right)}{\sinh \left(x - i \left(\frac{\pi}{4}-\frac{(2n+1)\gamma}{4}\right)\right)}
\right)^L
\!\!\!\!\! &=& \!\!\!\!\!
\prod_{x'}  \frac{\sinh\left(x-x' - i \gamma \right)}{\sinh\left( x-x' \right)} 
\frac{\sinh \left(x-x'  - i \left(\frac{\pi}{2}-\frac{2n-1}{2}\gamma\right)\right)}{\sinh \left(x-x' - i \left(\frac{\pi}{2}-\frac{2n+1}{2}\gamma\right)\right)} \,,
\end{eqnarray}
while the rest of the Bethe equations become trivial. 
Multiplying the two above relations one gets 
\begin{align}
& &
\left(  
\frac{\sinh \left(x + i \left( \frac{\pi}{4}-\frac{(2n+1)\gamma}{4}\right)\right)}{\sinh \left(x - i \left(\frac{\pi}{4}-\frac{(2n+1)\gamma}{4}\right)\right)}
\frac{\sinh \left(x - i \left(\frac{\pi}{4}+\frac{(2n-3)\gamma}{4}\right)\right)}{\sinh \left(x + i \left(\frac{\pi}{4}+\frac{(2n-3)\gamma}{4}\right)\right)}
\right)^L
\nonumber \\
& & =
\prod_{x'} 
\frac{\sinh \left(x-x'-i \gamma\right)}{\sinh \left(x-x' + i \gamma\right)} 
\frac{\sinh \left(x-x'  + i \left(\frac{\pi}{2}-\frac{2n+1}{2}\gamma\right)\right)}{\sinh \left(x-x'  - i \left(\frac{\pi}{2}-\frac{2n+1}{2}\gamma\right)\right)}
\frac{\sinh \left(x-x'  - i \left(\frac{\pi}{2}-\frac{2n-1}{2}\gamma\right)\right)} 
{\sinh \left(x-x'  + i \left(\frac{\pi}{2}-\frac{2n-1}{2}\gamma\right)\right)} \,.
\label{eq:ane2RII_BAEreal}
\end{align} 

In Fourier space this becomes 
\begin{eqnarray}
\rho + \rho^h &=& \frac{\sinh \omega \left(\frac{\pi}{4} + \frac{(2n-3)\gamma}{4}\right)+\sinh \omega \left(\frac{3\pi}{4} - \frac{(2n+1) \gamma}{4}\right)}{\sinh \frac{\omega\pi}{2}}
- \nonumber \\
& & \rho \,
\frac{\sinh ({\pi}- \frac{2n+1}{2}\gamma)\omega \gamma + \sinh \frac{2n-1}{2}\omega \gamma+\sinh \omega \left(\frac{\pi}{2} -\gamma\right)}{\sinh \frac{\omega\pi}{2}} \,. \label{rhorhoha2n2}
\end{eqnarray}
The matrix $K$ is now a simple scalar, and its zero frequency limit is 
\begin{equation}
 1- K(0) = 4\left(1  -  \frac{\gamma}{\pi}\right) \,.
\end{equation}

Note however that many more excitations are possible than creating holes of complexes (in particular, the complexes can be partly broken). Numerical study shows that the central charge is $c=n+{1\over 2}$, suggesting the presence of an additional Majorana fermion.

Staggering like in regime I gives results compatible with a Toda theory coupled to a Majorana fermion, with  action  
\begin{equation}
S=\int {1\over 2} (\partial_\mu\pmb{\phi} \cdot \partial_\mu\pmb{\phi})+ \psi\partial_\mu\psi+\bar{\psi}\partial_\mu\bar{\psi}+g\left[e^{-2i\beta\phi_1}+2\sum_{i=1}^{n-1}e^{i\beta(\phi_i-\phi_{i+1})}+\bar{\psi}\psi e^{i\beta\phi_{n}}\right] \,.
\label{Todageni3}\end{equation}
This is in fact  an imaginary version of the  super algebra $B^{(1)}(0,n)$ Toda theory \cite{Olsh}. The mass ratios in this theory are in general coupling dependent.

\subsection{The case $a_{2n-1}^{(2)}$ in regime II}
\label{sec:ano2RII}

This corresponds to ${\cal N}>0$ and $\gamma\in [{\pi\over 2n},{\pi\over 2}]$.
From explicit study of the cases $a_{3}^{(2)}$ and $a_{5}^{(2)}$, we conjecture that the ground state in this regime is described now by the following patterns of roots
\begin{eqnarray}
\lambda^{\alpha} &\simeq & x \pm i \left(\frac{\pi}{4} - \frac{(2(n-\alpha)) \gamma }{4}\right) \,, \qquad \mbox{for } \alpha = 1,2,\ldots,n-1 \,, \nonumber \\
\lambda^{n} &=& x + i\frac{\pi}{4} \,.
\end{eqnarray}

The same manoeuvre as in the $a_{2n}^{(2)}$ case gives equations the for densities, which read in Fourier space
\begin{eqnarray}
\rho + \rho^h &=& \frac{\sinh \omega \left(\frac{\pi}{4} + \frac{(n-2)\gamma}{2}\right)+\sinh \omega \left(\frac{3\pi}{4} - \frac{n \gamma}{2}\right)}{\sinh \frac{\omega\pi}{2}}
- \nonumber \\
& & \rho 
\frac{\sinh ({\pi}- n\gamma)\omega \gamma + \sinh (n-1)\omega \gamma+\sinh \omega \left(\frac{\pi}{2} -\gamma\right)}{\sinh \frac{\omega\pi}{2}} \,,
\end{eqnarray}
The matrix $K$ is again a scalar, and its zero frequency limit is 
\begin{equation}
 1- K(0) = 4\left(1  -  \frac{\gamma}{\pi}\right) \,.
\end{equation}

There is strong evidence that properties of this regime (including the massive deformation produced by staggering)  are the continuation of those in regime I,  the relationship between
the conformal weight  of the perturbation in (\ref{Todagenii}) and $\gamma$ becoming then
\begin{equation}
{\beta^2\over 8\pi} ={\pi-\gamma\over 2\pi}
\end{equation}
instead of (\ref{dimcond}). The central charge is $c=n$. The constant $g$ becomes dimensionless when 
\begin{equation}
{\pi-\gamma\over 2\pi}={2n-1\over 4n} \quad \Leftrightarrow \quad \gamma={\pi\over 2n} \,,
\end{equation}
which corresponds exactly to the junction of regimes I and II. 

\subsection{Example 1: $a_2^{(2)}$}
\label{sec:a22RII}

The case  ${\cal N}>0$ and $\gamma \in \left[ \frac{\pi}{3},\pi \right]$ (resp.\ $\gamma \in \left[ 0,\frac{\pi}{3} \right]$) corresponds to the regimes called II (resp.\ III) in \cite{Nienhuis}, and is much more difficult to analyse than regime I. A naive analysis would suggest that the ground state is obtained by filling a sea of  real $\lambda_j$'s, but this is not the case. In fact, the ground state is made of complexes with imaginary parts close to $\pm {1\over 4}(\pi-\gamma)$:
\begin{equation}
\lambda_j= x_j \pm {i\over 4}(\pi-\gamma) \,.
\end{equation}
Note that these two-strings are not the usual ones, since the gap in imaginary parts is equal to ${1\over 2}(\pi-\gamma)$ rather than $\gamma$; this is possible because the right-hand side of the Bethe equations (\ref{BetheEqs2}) contains a ratio of cosine terms, that results from the twisting of $a_2$. 

The same two-strings build the ground state in regime II and regime III. Differences arise however in the corrections to the asymptotic shape of the complexes, as well as the  analytical behaviour of the Bethe kernels. We discuss here regime II, which  corresponds to $\gamma\in [{\pi\over 3},\pi]$.

The bare equations in this regime read
\begin{eqnarray}
\rho+\rho^h &=& {2\sinh {\omega\over 2}(\pi-\gamma)\cosh {\omega\over 4}(\pi-\gamma)\over \sinh {\omega\pi\over 2}}- \nonumber \\
& & {\sinh \omega\left({\pi\over 2}-\gamma\right)-\sinh {\omega\over 2}(3\gamma-2\pi)+\sinh {\omega\gamma\over 2}\over \sinh {\omega\pi\over 2}}\rho \,.
\end{eqnarray}
Thus the matrix $K$ at zero frequency is the scalar $K={4\gamma\over\pi}-3$, so $1-K=4{\pi-\gamma\over\pi}$, which leads to the  spectrum of conformal weights associated with the formation of holes (recall that we do  not discuss the effects of shifts of the sea)
\begin{equation}
\Delta+\bar{\Delta}={\pi-\gamma\over \pi}n_h^2 \,, \label{fssII}
\end{equation}
where $n_h$ is the number of holes of complexes. Since this  corresponds to removing two Bethe roots, we have as well 
\begin{equation}
\Delta+\bar{\Delta}={\pi-\gamma\over 4\pi}(\delta m)^2 \,. \label{fssII}
\end{equation}
The central charge is found to be $c={3\over 2}$, and  there are now more excitations, which can be identified with the presence of a Majorana fermion. 

The physical equations  are 
\begin{equation}
\rho+\rho^h=s+K \star\rho^h \,,
\end{equation}
where, in Fourier variables, %
\begin{eqnarray}
s&=&{1\over 2\cosh{\omega\over 4}(3\gamma-\pi)} \,, \nonumber \\
K &=& {1\over 4\cosh^2{\omega\over 4}(\pi-\gamma)}-{\sinh{\omega\over 2}(3\gamma-2\pi)\over 4\sinh {\omega\over 2}(\pi-\gamma)\cosh{\omega\over 4}(\pi-\gamma) \cosh{\omega\over 4}(3\gamma-\pi)} \,. \label{scatt}
\end{eqnarray}

Staggering the bare spectral parameter so as to interpret our theory as the UV limit of a massive integrable QFT leads to most interesting results. First, we observe that the $S$-matrix which appears in (\ref{scatt}) has not, to the best of our knowledge, appeared in the literature before. 

With staggering determined by   $\Lambda$, we find, from the pole at 
$\omega={2i\pi\over 3\gamma-\pi}$ in the Fourier integral for the source term, that the mass scale induced is 
\begin{equation}
M\propto \exp\left[-\Lambda~ {2\over 3{\gamma\over\pi}-1}\right] \,.
\end{equation}
%
%
%
But we know from our earlier study (\ref{couprel}) that $g\propto e^{-{4\over 3}\Lambda}$. This implies that the bare coupling obeys
\begin{equation}
g\propto [\hbox{length}]^{{2\over 3}-{2\gamma\over \pi}} \,. \label{coupreli}
\end{equation}
We claim that this corresponds to  the following perturbation:
\begin{equation}
S=\int {1\over 2} (\partial_\mu\phi)^2+ \psi\partial_\mu\psi+\bar{\psi}\partial_\mu\bar{\psi}+g\left[e^{-2i\beta\phi}+\psi\bar{\psi}e^{i\beta\phi}\right] \,, \label{ourpert}
\end{equation}
which is usually referred to as the $B^{(1)}(0,1)$ Toda theory \cite{Olsh,MathieuWatts}. Indeed,   we see that  for this action we need to have
\begin{equation}
[g]^3 \left([\hbox{length}]^{-1+(\beta)^{2}/4\pi}\right)^2[\hbox{length}]^{-2+(\beta)^2/\pi}\propto 1 \,,
\end{equation}
so 
\begin{equation}
[g]=[\hbox{length}]^{-{4\over 3}+{(\beta)^2\over 2\pi}} \,,
\end{equation}
and this matches (\ref{coupreli}) provided that 
\begin{equation}
{(\beta)^2\over 8\pi}={\pi-\gamma\over2\pi} \,.
\end{equation}
Of course, this dimension is allowed by the finite-size spectrum (\ref{fssII}). The extra ($\beta$ independent) dimensions associated with the fermionic degrees of freedom $\bar{\psi},\psi$ would appear, as usual, following a more complete analysis of the finite-size effects in the presence of strings \cite{Nienhuis}. The coupling becomes dimensionless at the edge of the regime, here for $\gamma={\pi\over 3}$. 
\bigskip

We note here that, apart form the $B^{(1)}(0,1)$ theory, there are two other integrable Toda theories involving one boson and one Majorana fermion:
\begin{enumerate}
 \item The $C^{(2)}(2)$ theory:
\begin{equation}
S=S_{\rm FBc}+S_{\rm Ising}+g\int {\rm d}^2z \, \psi\bar{\psi}\cos\beta\phi \,.
\end{equation}
 \item The $A^{(4)}(0,2)$ theory:
\begin{equation}
S=S_{\rm FBc}+S_{\rm Ising}+g\int {\rm d}^2z \, \left[\psi\bar{\psi}e^{-i\beta\phi}+e^{i\beta\phi}\right] \,,
\end{equation}
\end{enumerate}
where $S_{\rm FBc}$ and $S_{\rm Ising}$ denote the actions for a free compact boson and a free Majorana fermion, respectively.
These two other possibilities can however be discarded by a careful analysis of our equations. Observe also that, for the theory in (\ref{ourpert}), there are 
 two non-local conserved currents for our theory, one fermionic and one bosonic:
\begin{eqnarray}
J_1&=&\psi e^{-{4i\pi\over \beta}\phi_{\rm R}} \,, \nonumber\\
J_2&=&e^{{4i\pi\over\beta}\varphi} \,,
\end{eqnarray}
where $\phi_{\rm R}$ is the right-moving part of the field $\phi$. Note that the theory perturbed by the currents
\begin{equation}
S=S_{\rm FB}+S_{\rm Ising}+g\int {\rm d}^2z \, \left[\psi\bar{\psi}e^{-{4i\pi\over \beta}\phi}+e^{{4i\pi\over\beta}\phi}\right]
\end{equation}
has exactly the form of an $A^{(4)}(0,2)$ theory, so the two types are obviously dual of each other. 
Finally, note that   the $a_2^{(2)}$ model at $\gamma={\pi\over 2}$ is equivalent to the antiferromagnetic Fateev-Zamolodchikov model. The latter  model is obtained via the $so(3)^{(1)}$ solution of the Yang-Baxter equation, which coincides with  the spin-one $su(2)$ solution. In that case, the $B^{(1)}(0,1)$ theory and the $C^{(2)}(2)$ theories are equivalent. 

\bigskip

To further justify our identification of the continuum limit, we can now explore the scattering theory in more details. If we believe indeed that the $S$-matrix describes a complex version of the $B^{(1)}(0,1)$ Toda theory, some of the  results which are known  for this theory at real (also often called `physical', since then the action is real) coupling carry over to the complex case. This is addressed briefly in 
 \cite{Baseilhac}, where   we set $\xi={\pi\over 3\pi-4\gamma}$. It follows that the relation between the lightest breather mass $\overline{m}$ and the kink mass $M$ for the theory (\ref{ourpert}) should be, if the usual relationship between the theories at real and imaginary coupling holds, 
\begin{equation}
\overline{m}=2M\sin\pi {\pi(1+\xi)\over 5\xi-3}=2M\sin{\pi\over 2T-3} \,,
\end{equation}
where we have set, to make our notations lighter,
\begin{equation}
\gamma=\pi-{\pi\over T} \,.
\end{equation}
To get this, we start from the real Toda theory \cite{BasFat} at coupling $b^2$. Consider the fundamental fermion $S$-matrix, which reads, in the real Toda theory
\begin{equation}
S_{\psi\psi}={\sinh \theta+i\sin{2\pi\over h}\over \sinh\theta-i\sin{2\pi\over h}}\times{\sinh\theta-i\sin\left({2\pi\over h}-{\pi\Omega\over h}\right)\over
\sin\theta+i\sin\left({2\pi\over h}-{\pi\Omega\over h}\right)} \,,
\end{equation}
where we have
\begin{eqnarray}
h&=&{2+3b^2\over 1+b^2} \,, \nonumber \\
\Omega&=&{b^2\over 1+b^2} \,.
\end{eqnarray}
Replacing $b^2$ by $-{(\beta')^2\over 8\pi}$ yields the analytically continued first breather $S$-matrix
\begin{equation}
S_{\psi\psi}\to{\sinh \theta-i\sin{\pi\over 2T-3}\over \sinh\theta+i\sin{\pi\over 2T-3}}\times{\sinh\theta+i\sin{2\pi\over 2T-3}\over
\sin\theta-i\sin{2\pi\over 2T-3}} \,.
\end{equation}
In Fourier variables this becomes
\begin{equation}
{1\over i}{{\rm d} \over {\rm d} \theta}\ln S=\int {\rm d}\kappa \, e^{-i\kappa\theta} {\cosh {\kappa \pi(2T-5)\over 2(2T-3)}-\cosh{\kappa \pi(2T-7)\over 2(2T-3)}\over \cosh {\kappa\pi\over 2}} \,.
\end{equation}
We can now go back to our Bethe equations. In terms of the $z$-variables, the ground state is formed of two-strings $z=\xi\pm {i\Gamma\over 2}$, and the most natural other excitations to consider are antistrings, $z=w+i\pi$. Calling the density of these  excitations $\rho_1$, one finds after a few manipulations the Bethe equations
\begin{eqnarray}
\rho_1+\rho_1^h &=& {\cosh{\omega\over 4}(5\gamma-3\pi)\over \cosh{\omega\over 4}(3\gamma-\pi)}+{\cosh{\omega\over 4}(5\gamma-3\pi)-\cosh{\omega\over 4}(7\gamma-5\pi)\over \cosh{\omega\over 4}(3\gamma-\pi)}\rho_1 \nonumber \\
& & -{\cosh{3\omega\over 2}(\pi-\gamma)\over 2\cosh{\omega\over 4}(3\gamma-\pi)\cosh{\omega\over 4}(\pi-\gamma)}\rho^h \,,
\end{eqnarray}
where $\rho^h$ is the density of holes in the Fermi sea. This is in complete agreement with the foregoing identification and $\kappa=\omega {3\gamma-\pi\over2\pi}$. In particular, we see that the mass of the bound state is 
\begin{equation}
\bar{m}=2M\cos(5\gamma-3\pi){\pi\over 2(3\gamma-\pi)}=2\sin{\pi\over 2T-3}
\end{equation}
indeed. 

It is possible to  build the whole scattering theory using these ingredients. This is however not our purpose here, so we will cut the discussion short, and content ourselves with the conclusion that the identification (\ref{ourpert}) is the correct one.

\subsection{Example 2: $a_3^{(2)}$}
\label{sec:a32RII}

In this regime, which extends over $\gamma\in [{\pi\over 4},{\pi\over 2}]$, we find  that the ground state is made of strings over strings, in the form
\begin{eqnarray}
\lambda^1&=&x\pm i\left({\pi\over 4}-{\gamma\over 2}\right) \,, \nonumber\\
\lambda^2&=&x+{i\pi\over 4} \,.
\end{eqnarray}
The bare 
equations are now
\begin{equation}
-(\rho+\rho^h)=-{\sinh \omega{\pi\over 4}+\sinh \omega({3\pi\over 4}-\gamma)\over \sinh {\omega\pi\over 2}}+\left({\sinh \omega\gamma+\sinh \omega(\pi-2\gamma)+\sinh \omega({\pi\over 2}-\gamma)\over \sinh {\omega\pi\over 2}}\right)\rho \,.
\end{equation}
This leads to $K={4\gamma\over \pi}-3$, and a spectrum due to holes
\begin{equation}
\Delta+\bar{\Delta}=\left(1-{\gamma\over\pi}\right)n_h^2 \,.
\end{equation}
The physical equations are 
\begin{equation}
\rho={1\over 2\cosh \omega({\pi\over 4}-\gamma)}-{\sinh {\omega\pi\over 2}\over 4\sinh \omega({\pi\over 2}-{\gamma\over 2}) \cosh \omega({\pi\over 4}-{\gamma\over 2}) \cosh \omega({\pi\over 4}-\gamma)}\rho^h \,.
\end{equation}

A more detailed analysis suggests that this regime is in fact the continuation of regime I beyond ${\pi\over 2}$. For instance, under the usual staggering and using the pole at $\omega={i\pi\over 2}{1\over \gamma-{\pi\over 4}}$, we have
\begin{equation}
[g]=[\hbox{length}]^{-2+{8\over 3}(1-{\gamma\over \pi})} \,,
\end{equation}
which is exactly the continuation of the equation in regime I after substitution $\gamma\to \pi-\gamma$.

\subsection{Example 3: $a_4^{(2)}$}
\label{sec:a42RII}

\begin{figure}
\begin{center}
 \includegraphics[scale=0.8]{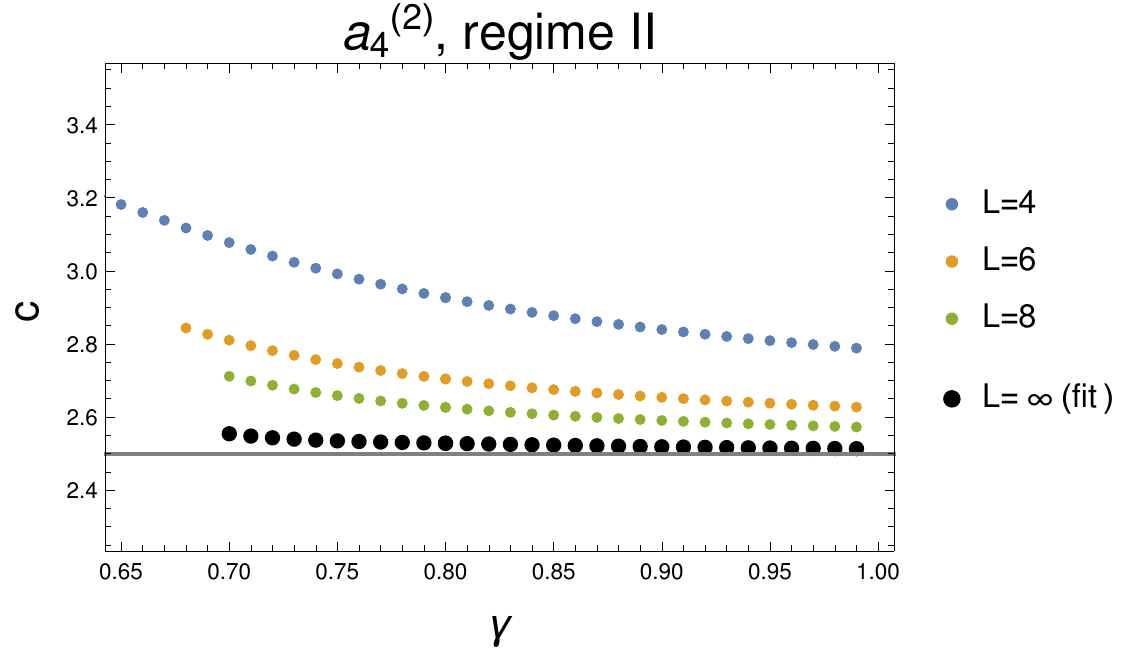}
 \end{center}
\caption{Measure of the central charge in the regime II of the $a_4^{(2)}$ model, from direct diagonalisation of the transfer matrix. The black dots represent an extrapolation to $L\to \infty$, using a quadratic fit in $1/L$.}                                 
\label{fig:c_a42RII}                          
\end{figure}

For $\gamma < {\pi \over 3}$, the roots have the generic structure reported in section \ref{sec:ane2RII}, namely 
 \begin{eqnarray}
\lambda^1 &\simeq& x \pm i \left(\frac{\pi}{4} - \frac{3\gamma}{4}\right) \,, \nonumber \\
\lambda^2 &\simeq& x \pm i \left(\frac{\pi}{4} - \frac{\gamma}{4}\right) \,,
 \end{eqnarray}  
 leading to scattering equations which have the same form as those discussed in section \ref{sec:ane2RII}.
The measure of central charge is given in figure \ref{fig:c_a42RII}, leading to the conjecture $c=\frac{5}{2}$.

Now turn to $\gamma > {\pi \over 3}$. 
At ${\pi \over 3}$ the $\lambda^1$ two-strings have zero imaginary part. We observed that past this value the $\lambda^1$-roots lie on the real axis. 
In finite size the transition between these two regimes does not happen exactly at $\gamma = {\pi \over 3}$, but in a really narrow region around it: the $\lambda^1$ 2-strings in the centre of the Fermi sea have smaller imaginary part and become real as the others are well separated. 
The physical equations now involve two different densities of real parts,  $\rho_1$ and $\rho_2$ respectively, and read
\begin{eqnarray}
 \rho_1 &=&  \frac{\cosh{\omega \over 4}(\pi - 3 \gamma)}{\cosh{\omega \over 4}(\pi - 5 \gamma)} -  \frac{\cosh{\omega \over 4}(\pi - 3 \gamma)}{\cosh{\omega \over 4}(\pi - 5 \gamma)}\frac{\sinh \omega {\pi \over 2}}{\sinh \omega \left({\pi \over 2}-{\gamma \over 2}\right)}\rho_1^h \nonumber \\
 & & -  \frac{1}{\cosh{\omega \over 4}(\pi - 5 \gamma)}\frac{\sinh \omega {\pi \over 2}}{\sinh \omega \left({\pi \over 2}-{\gamma \over 2}\right)}\rho_2^h \,, \\
  \rho_2 &=&  \frac{\cosh{\omega \over 4}(\pi - 3 \gamma)}{\cosh({\omega \over 4}(\pi - 5 \gamma)} -  \frac{1}{2\cosh{\omega \over 4}(\pi - 5 \gamma)}\frac{\sinh \omega {\pi \over 2}}{\sinh \omega\left( {\pi \over 2}-{\gamma \over 2}\right)}\rho_1^h \nonumber \\
 & & -  \frac{\cosh \omega{\gamma \over 2}}{\cosh{\omega \over 4}(\pi - 5 \gamma)\cosh \omega \left({\pi \over4}-{\gamma \over 4}\right)}\frac{\sinh \omega {\pi \over 2}}{\sinh \omega \left({\pi \over 2}-{\gamma \over 2}\right)}\rho_2^h \,.
\end{eqnarray}
Hence the ground state distributions
\begin{eqnarray}
\rho_1 &=&  \frac{\cosh {\omega \over 4}(\pi - 3 \gamma)}{\cosh{\omega \over 4}(\pi - 5 \gamma)} \,, \\
\rho_2 &=& \frac{1}{\cosh{\omega \over 4}(\pi - 5 \gamma)} \,.
\end{eqnarray}
This leads to an integral expression of the ground state energy in the thermodynamic limit, which turns out to be the analytical continuation of that in the 2-string case. The Fermi velocity is also the same, and numerical measures of the central charge indicate that $c=\frac{5}{2}$ holds all through regime II. From there, it seems reasonable to conjecture that all conformal properties are unchanged as $\gamma$ is varied between $\frac{\pi}{5}$ and $\frac{\pi}{2}$.

\subsection{General comments}

It is not clear to us physically why  the $a_{2n}^{(2)}$ models exhibit a regime where the continuum limit involves fermions, while the $a_{2n-1}^{(2)}$ models do not. From a lattice point of view, the fermions seem to have something to do with the states in the $\check{R}$ matrix carrying vanishing spin: these occur only for $a_{2n}^{(2)}$, since then the fundamental representation has an odd ($2n+1$) number of sites. 

\section{Regime III}
\label{sec:III}

This is the most interesting of all the regimes, where we claim that the continuum limit systematically involves non-compact degrees of freedom. 

We have gathered experience on this regime with our earlier studies of the $a_2^{(2)}$ and $a_3^{(2)}$ cases \cite{VJS1,VJS2}. Our expectation, based on these studies and a strong duality argument (see below) is that the continuum limit is given by a system of compact and non-compact bosons which can be seen as the natural Coulomb gas representation of the $SO(N)/SO(N-1)$ cosets for $a_{N-1}^{(2)}$. The expected central charge and types of bosons are discussed below. For convenience, a summary of our findings is given in table~\ref{summ}. 

\begin{table}
\begin{center}
$$
\begin{array}{l|l|l|l}
 &\text{Regime I} & \text{Regime II} & \text{Regime III} \\
 \hline
 a_{2n}^{(2)}  & \gamma \in \big[0,\pi \big] \mbox{and } {\cal N }<0   & \gamma \in \big[{\pi \over 2n+1},\pi \big] \mbox{and } {\cal N }>0 & \gamma \in \big[ 0,{\pi \over 2n+1} \big] \mbox{and } {\cal N }>0  \\[2mm]
 & c=n  & c=n+\frac{1}{2} & c=2n \\[2mm]
  & \mbox{$n$ compact bosons}
  & \left \lbrace \begin{array}{ll} \mbox{$n$ compact bosons}  \\ \mbox{1 Majorana fermion} \end{array} \right.
  & \left\lbrace \begin{array}{ll} \mbox{$n$ compact bosons} \\ \mbox{$n$ non-compact bosons} \end{array} \right.  \\[3mm]
  & \mbox{$a_{2n}^{(2)}$ Toda} & \mbox{$B^{(1)}(0,n)$ Toda}  &  \\[1mm]
  \hline
   a_{2n-1}^{(2)}  & \gamma \in \big[ 0,\pi \big] \mbox{and } {\cal N }<0 & \gamma \in \big[{\pi \over 2n},\pi \big] \mbox{and } {\cal N }>0  &  \gamma \in \big[0,{\pi \over 2n} \big] \mbox{and } {\cal N }>0 \\[2mm]
 & c=n  & c=n & c=2n-1 \\[2mm]
  & \mbox{$n$ compact bosons} & \mbox{$n$ compact bosons}
  & \left \lbrace \begin{array}{ll} \mbox{$n$ compact bosons} \\ \mbox{$n-1$ non-compact bosons} \end{array} \right. \\[3mm]
  & \mbox{$a_{2n-1}^{(2)}$ Toda} & \mbox{$a_{2n-1}^{(2)}$ Toda}   &   \\
\end{array}
$$
\caption{Summary of our results.}
\label{summ}
\end{center}
\end{table}

\subsection{Duality}
\label{sec:duality}

To explain the duality argument, we  go back to the general relations satisfied by the algebra generators; see eqs.~(\ref{skein}) and (\ref{braidrel})--(\ref{tanglerel}). It is easy  to prove algebraically from them that, abstractly, the two $\check{R}$-matrices, (\ref{R1def}) and (\ref{newsol}), satisfy identical algebraic relations for matching values of the parameters:
\begin{equation}
 \check{R}^{(1)}, \ \check{R}^{(2)}, \ SO(N), \ q, \ x \quad \longleftrightarrow
 \quad \check{R}^{(2)}, \ \check{R}^{(1)}, \ SO(\widetilde{N}), \ q, \ x^{-1} \,, \label{equivalence}
\end{equation}
where we have set $q = \exp \left( \frac{i \pi}{N + \widetilde{N} - 2} \right)$.
This is done by comparing the relations satisfied by the generators, and using the fact that $q^{N-1}=-\left(q^{-1}\right)^{\widetilde{N}-1}$.  

Of course, algebraic equivalence is not the end of the story. First of all, objects such as Birman-Wenzl generators can satisfy identical relations but not be identical because they correspond, in technical terms, to different {\sl representations} of the algebra. Moreover, the full argument is based on $\check{R}$-matrices. These give rise to vertex models with `twists',%
\footnote{The use of `twisted' in this context refers to the boundary conditions of the lattice model, or the addition of a charge at infinity for the field theory. It is not related with the fact that the Lie algebras underlying the $a_{N-1}^{(2)}$ model are twisted.}
whose properties can be different from these without twists. In some cases, this difference is easily taken into account by changing the boundary conditions in the same continuum limit theory. In other cases, the difference is more profound, and can lead to different universality classes. 

A more thorough analysis of the meaning of the equivalence (\ref{equivalence}) is possible, along the lines of the level-rank duality analysis in the $SU(N)$ case \cite{AltSal}. The result is that one expects full coincidence of the truncated, RSOS versions. 

Now, although the continuum limit of the $so(\widetilde{N})$ RSOS models (i.e. those associated with $\check{R}^{(1)}$) is not entirely understood, it is believed that there is a regime where it is simply given by diagonal GKO cosets \cite{GKO}
\begin{equation}
{SO(\widetilde{N})_1\times SO(\widetilde{N})_l\over SO(\widetilde{N})_{l+1}} \,,
\end{equation}
with the level easily related 
to the quantum group deformation parameter
\begin{equation}
q=\exp \left({i\pi \over l+\widetilde{N}-1} \right).
\end{equation}
It is well-known that these CFTs can also be formulated  as different cosets. This can be seen for instance by studying the central charge
\begin{eqnarray}
{SO(\widetilde{N})_1\times SO(\widetilde{N})_l\over SO(\widetilde{N})_{l+1}} \,: \quad c={\widetilde{N}\over 2}{l(l+2\widetilde{N}-3)\over (\widetilde{N}+l-1)(\widetilde{N}+l-2)} \,, \nonumber\\
{SO(N)_k\over SO(N-1)_k} \,: \quad c={k\over 2} {(N-1)(2k+N-4)\over (k+N-2)(k+N-3)} \,,
\end{eqnarray}
and checking that the two coincide for $l=N-1$ and $k=\widetilde{N}$. In fact, there is a full conformal 
duality \cite{Fuchs}
\begin{equation}
{SO(\widetilde{N})_1\times SO(\widetilde{N})_{N-1}\over SO(\widetilde{N})_{N}} \quad \longleftrightarrow \quad {SO(N)_{\widetilde{N}}\over SO(N-1)_{\widetilde{N}}} \,.
\end{equation}
Putting together the conformal and lattice algebraic duality, we conclude that  there is a regime where the continuum limit of the model given by $\check{R}^{(2)}$ is the $ SO(N)_{\widetilde{N}} / SO(N-1)_{\widetilde{N}}$ coset model, where   $q=e^{i\pi/(N+\widetilde{N}-2)}$. Moreover, since $\widetilde{N}\geq 2$ for these equations to make sense, it is reasonable to expect that the corresponding regime covers $\gamma\in [0,{\pi\over N}]$: in particular, this means $\gamma\in [0,{\pi\over 2n}]$ for $a_{2n-1}^{(2)}$, which is associated with $so(N=2n)$. Of course, when $\widetilde{N}$ is not an integer, the argument per se does not apply. Previous experience with the $a_2^{(2)}$ case \cite{VJS1} shows however that the argument extends to the case of real $\widetilde{N}$, provided the coset models are replaced by the appropriate Coulomb gas, and, for the lattice vertex model, the charge at infinity is set to zero. 

Some features follow immediately from the detailed discussion given in the appendix. We see in particular that we have the pattern:
$$
\begin{array}{l|l|l|l}
a_{2n-1}^{(2)} & \hbox{Rank $n$} &
\left \lbrace \begin{array}{ll} \mbox{$n$ compact bosons} \\ \mbox{$n-1$ non-compact bosons} \end{array} \right.
& c=2n-1\\ \hline
a_{2n}^{(2)} & \hbox{Rank $n$} & \left\lbrace \begin{array}{ll} \mbox{$n$ compact bosons} \\ \mbox{$n$ non-compact bosons} \end{array} \right.
& c=2n
\end{array}
$$
The number of compact bosons is the same as the one we have observed in regime I.

We can go one step further and discuss also the effect of staggering. In general, the staggering of the $SO(\tilde{N})$ vertex model will correspond, in the twisted theory
\begin{equation}
 {SO(\widetilde{N})_1\times SO(\widetilde{N})_{N-1}\over SO(\widetilde{N})_{N}} \,,
\end{equation}
to a perturbation with conformal weight (see also \cite{Vaysburd})
\begin{equation}
\Delta=1-{\widetilde{N}-2\over \widetilde{N}+N-2}={N\over  \widetilde{N}+N-2} \,,
\end{equation}
so that, e.g.\ for $N=3$, we have indeed
\begin{equation}
 h={3\over \widetilde{N}+1}={6\over 2\widetilde{N}+2} \,,
\end{equation}
which is the dimension of the second energy operator for the $Z_{2\widetilde{N}}$ model.

Let us now turn to the results of the lattice model analysis.

\subsection{Root patterns and (some features of) the compact sector}

We recall that regime III corresponds to ${\cal N}>0$ and $\gamma\in [0,{\pi\over 2n}]$ for $a_{2n-1}^{(2)}$, resp. $\gamma\in [0,{\pi\over 2n+1}]$ for $a_{2n}^{(2)}$. 
The roots patterns for the ground state were found (explicitly for $a_2^{(2)},a_3^{(2)},a_4^{(2)},a_5^{(2)}$) to have a similar form as those corresponding to regime II, namely:
\begin{enumerate}
 \item For $a_{2n-1}^{(2)}$:
\begin{eqnarray}
\lambda^{\alpha} &\simeq&  x \pm i \left(\frac{\pi}{4} - \frac{(2(n-\alpha)) \gamma }{4}\right) \,, \qquad \mbox{for } \alpha = 1,2,\ldots,n-1
\nonumber \\
\lambda^{n} &=& x + i\frac{\pi}{4} \,.
\end{eqnarray}
 \item For $a_{2n}^{(2)}$:
\begin{equation}
\lambda^{\alpha} \simeq  x \pm i \left(\frac{\pi}{4} - \frac{(2(n-\alpha)+1) \gamma }{4}\right)\,, \qquad \mbox{for } \alpha = 1,2,\ldots,n \,.
\end{equation}
\end{enumerate}
The only qualitative difference resides in the sign of the corrections to the real and imaginary parts of the various 2-strings. 
The form of the Bethe equations in real space is therefore the same as in regime II, namely (\ref{eq:ane2RII_BAEreal}) for $a_{2n}^{(2)}$, and its counterpart for $a_{2n-1}^{(2)}$ respectively. However, the determinations of the logarithms are then different, and the continuous Bethe equations in Fourier space take a different form, namely:
\begin{enumerate}
 \item For $a_{2n-1}^{(2)}$:
\begin{eqnarray}
\rho + \rho^h &=& \frac{\sinh \omega \left(\frac{\pi}{4} + \frac{n\gamma}{2}\right)-\sinh \omega \left(\frac{\pi}{4} + \frac{(n-2)\gamma}{2}\right)}{\sinh \frac{\omega\pi}{2}} \nonumber \\
& & - \rho 
\frac{\sinh n\omega \gamma - \sinh (n-1)\omega \gamma-\sinh \omega \left(\frac{\pi}{2} -\gamma\right)}{\sinh \frac{\omega\pi}{2}} \,,
\end{eqnarray}
leading to the following ground state ($\rho^h=0$) solution
\begin{equation}
\rho = \frac{1}{2 \cosh \frac{\omega}{4}(\pi-2n \gamma)} \,,
\end{equation}
which is the same as in regime II.
\item For $a_{2n}^{(2)}$:
\begin{eqnarray}
\rho + \rho^h &=& \frac{\sinh \omega \left(\frac{\pi}{4} + \frac{(2n+1)\gamma}{4}\right)-\sinh \omega \left(\frac{\pi}{4} + \frac{(2n-3)\gamma}{4}\right)}{\sinh \frac{\omega\pi}{2}} \nonumber \\
& & - \rho 
\frac{\sinh (n+\frac{1}{2})\omega \gamma - \sinh (n-\frac{1}{2})\omega \gamma-\sinh \omega \left(\frac{\pi}{2} -\gamma\right)}{\sinh \frac{\omega\pi}{2}}
\end{eqnarray}
[cf.~(\ref{rhorhoha2n2})], leading to the ground state solution, 
\begin{equation}
\rho = \frac{1}{2 \cosh \frac{\omega}{4}(\pi-(2n+1) \gamma)} \,,
\end{equation}
which once again has the same form as that of regime II.
\end{enumerate}
In both cases the matrix $K$ is a scalar, $K={4\gamma\over\pi}$. 

In regime III for $a_{2n}^{(2)}$, a hole corresponds to having  all the integers $\delta m_i=2$, so the conformal weight for $n_h$ holes of complexes is 
\begin{equation}
\Delta+\bar{\Delta}={\gamma\over \pi}(n_h)^2\equiv {\gamma\over\pi}(n_h)^2 \left( \sum_{i=1}^n\alpha_i\right)^2={\gamma\over\pi} (n_h)^2 \omega_1^2 \,,
\end{equation}
where we used that $\sum_{i=1}^n \alpha_i=e_1=\omega_1$ belongs to the $b_n$ weight lattice, so that $(\omega_i,\alpha^\vee_j)=\delta_{ij}$ with co-marks $\alpha^\vee_i={2\alpha_i\over |\alpha_i|^2}$. 

In regime III for $a_{2n-1}^{(2)}$, we have all the integers $\delta m_i=2$, except $\delta m_n=1$, so the conformal weight for $n_h$ holes of complexes is 
\begin{equation}
\Delta+\bar{\Delta}={\gamma\over 4\pi}(n_h)^2={\gamma\over 4\pi}\left( 2\sum_{i=1}^{n-1}\alpha_i+\alpha_n\right)^2={\gamma\over\pi} (n_h)^2 \omega_1^2 \,,
\end{equation}
where we used that $\sum_{i=1}^{n-1} \alpha_i+{1\over 2}\alpha_n=e_1$ again.
We note that $e_1\equiv \omega_1$ belongs to the weight lattice $(\omega_i,\alpha^\vee_j)=\delta_{ij}$, and $\alpha^\vee_i={2\alpha_i\over |\alpha_i|^2}$ of $c_n$ and also of  $d_n=so(2n)$. 

The identification of the continuum limit with a coset theory suggests that the whole spectrum of excitations in the compact sector should involve the norm square of vectors on the weight lattice of $so(2n+1)$ (resp. $so(2n)$) but we have not been able to check this. 
Indeed, unlike in regime I,  holes of complexes describe only a one-dimensional subset of the excitations in the compact sector. While an analysis of the other types of excitations necessary to understand this sector completely is in principle possible, it involves considerable technical difficulties, which are outside the scope of this paper. 

\subsection{The case $a_2^{(2)}$}
\label{sec:a22RIII}

Here regime III corresponds to  ${\cal N}>0$ and $\gamma\in [0,{\pi\over 3}]$.
It has  been discussed in great detail in our previous papers \cite{VJS1,VJS3}. The ground state is determined by the same complexes as in regime II, but the equations are changed due to analyticity properties of the kernels in Fourier space. One has now
\begin{equation}
\rho+\rho^h={2\sinh {\omega\gamma\over 2}\cosh \omega\left({\pi+\gamma\over 4}\right)\over \sinh {\omega\pi\over 2}}-{\sinh {3\omega\gamma\over 2}-\sinh {\omega\gamma\over 2}-\sinh \omega\left({\pi\over 2}-\gamma\right)\over \sinh {\omega\pi\over 2}}\rho \,,
\end{equation}
while the corresponding physical equations are
\begin{equation}
\rho={1\over 2\cosh {\omega\over 4}(\pi-3\gamma)}-{\sinh {\omega\pi\over 2}\over 
4\sinh {\omega\gamma\over 2}\cosh {\omega\over 4}(\pi+\gamma) \cosh {\omega\over 4}(\pi-3\gamma)}\rho^h \,.
\end{equation}

The central charge is found, after considerable analytical work, to be \cite{Nienhuis}
\begin{equation}
c=2 \,.
\end{equation}
Excitations obtained by removing complexes from the ground state can be handled analytically \cite{Nienhuis}. The final result is in agreement with the usual formula. 
We  have now, at zero frequency, $1-K=4{\gamma\over\pi}$, so the conformal weights associated with holes of complexes read
\begin{equation}
\Delta+\bar{\Delta}={\gamma\over 4\pi}(\delta m)^2 \,. \label{fssIII}
\end{equation}
Here, $\delta m$  is twice the number of holes of complexes; as a complex contains two Bethe roots, one has in fact $S^z=n$, with $S^z$ the spin of the excitation, in units where arrows in the vertex model carry $S^z=\pm 1$. 

Of course, since the central charge is $c=2$, there must be more degrees of freedom. The possibility of having two Majorana fermions---each contributing an extra ${1\over 2}$ to the central charge---is quickly excluded from numerics. There is, however, very strong evidence for a second bosonic degree of freedom, but a {\sl non-compact one}. This is discussed in great detail in our previous paper \cite{VJS1}. 

We can also investigate the---by now familiar---deformation obtained using staggering. The physical equations
give us the position of the poles and the mass scale as usual. Since we know the relationship between the staggering parameter $\Lambda$ and the bare coupling constant associated, we have now that 
\begin{equation}
[g]=[\hbox{length}]^{-{2\over 3}+{2\gamma\over\pi}} \,.
\end{equation}
A little exploration suggests that the associated theory corresponds to a perturbation of the form 
\begin{equation}
g\left[e^{-2i\beta\phi}+(1,1)e^{i\beta\phi}\right] \,, \label{explo}
\end{equation}
where $(1,1)$ denotes a field of weights $\Delta=\bar{\Delta}=1$ whose two-point function is non-zero. Matching dimensions gives
\begin{equation}
\lambda^3 L^6L^{-4}L^{-2\Delta}L^{-4\times \Delta/4}=1 \,, \qquad \Delta\equiv {\beta^2\over 2\pi} \,,
\end{equation}
so that
\begin{equation}
\lambda^3 L^{2-3\Delta}=1
\end{equation}
and thus 
\begin{equation}
{\beta^2\over 8\pi}={\gamma\over2\pi} \,,
\end{equation}
a value allowed by the finite-size spectrum (\ref{fssIII}). A more thorough study of this regime 
shows that the continuum limit can be described by two bosons $\phi,\varphi$, with 
$\varphi$ {\sl non-compact} and 
perturbation
\begin{equation}
g\left[e^{-2i\beta\phi}+\partial\varphi\bar{\partial}\varphi e^{i\beta\phi}\right] \,. \label{sick}
\end{equation}
The integrability of this theory can be formally established.%
\footnote{Note that action (\ref{sick}) presents unpleasant features, since one of the fields has dimension greater than two. Counter-terms are presumably necessary to make sense of the model.}
It can also be shown that it is related with the black hole sigma model and the $SU(2)/U(1)$ gauged WZW model, but we refrain from discussing this further here.

\subsection{The case $a_3^{(2)}$}
\label{sec:a32RIII}

Here regime III corresponds to ${\cal N}>0$ and $\gamma\in [0,{\pi\over 4}]$, and  has also been discussed in considerable detail in our previous work \cite{VJS2}.

In this regime, we find  that the ground state is made of strings over strings, in the form
\begin{eqnarray}
\lambda^1&=&x\pm i\left({\pi\over 4}-{\gamma\over 2}\right)+\epsilon_\pm \,, \nonumber\\
\lambda^2&=&x+{i\pi\over 4} \,,
\end{eqnarray}
where $\epsilon_\pm$ are infinitesimal quantities in the thermodynamic limit. 

%
%
%
%
%
%
%
%

\begin{figure}
\begin{center}
 \includegraphics[scale=0.8]{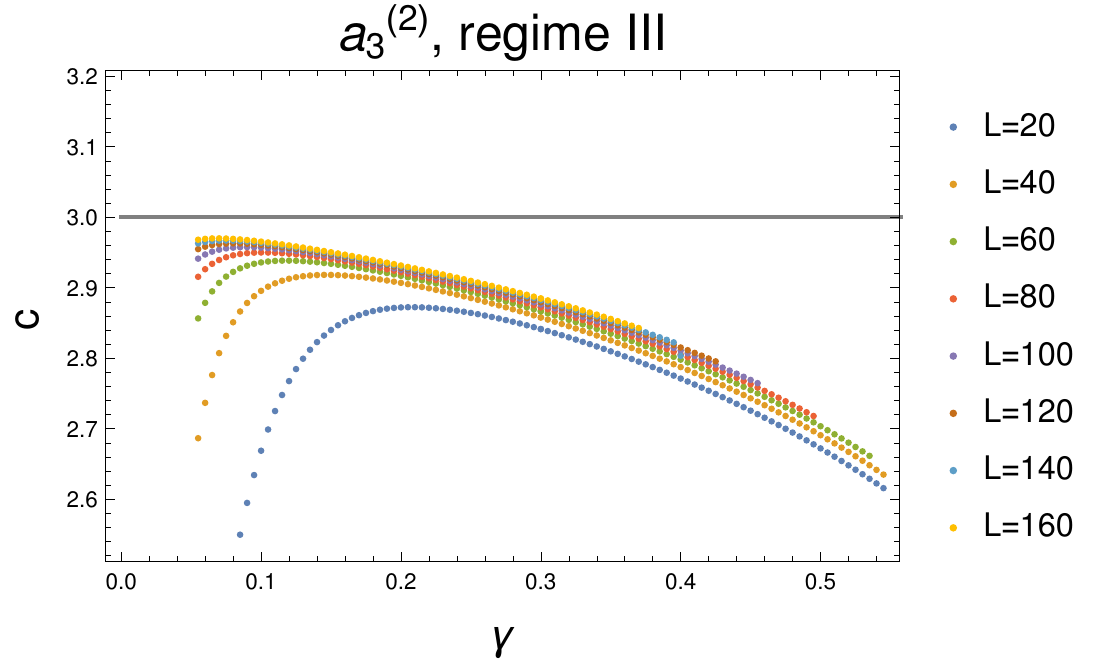}
 \end{center}
\caption{Measure of the central charge in regime III of the $a_3^{(2)}$ model, from numerical solution of the Bethe equations. Important finite-size corrections to the value $c=3$ are observed, associated with the non-compact continuum limit in this regime. }                           
\label{fig:c_a32RIII}                          
\end{figure}

Going over to Fourier transforms  ($\gamma<{\pi\over 4}$), and letting $\rho,\rho^h$ denote the densities of complexes and holes thereof, we find
\begin{equation}
\rho+\rho^h={\sinh \omega({\pi\over 4}+\gamma)-\sinh{\omega\pi\over 4}\over \sinh {\omega\pi\over 2}}+\left({\sinh \omega\gamma-\sinh 2\omega\gamma+\sinh \omega({\pi\over 2}-\gamma)\over\sinh{\omega\pi\over 2}}\right)\rho \,.
\end{equation}
The matrix $K$ is simply a scalar, $K=1-{4\gamma\over\pi}$, so we expect the hole contribution to the finite-size spectrum to be 
\begin{equation}
\Delta+\bar{\Delta}={\gamma\over \pi}n_h^2 \,.
\end{equation}
This should apply for $n_h$ holes in the ground state distribution. This means $2n_h$ $\lambda^1$-holes and $n_h$ $\lambda^2$-holes, so in terms of the magnetisations introduced in section \ref{sec:a32RI} we have $S_z=S_z'=n_h$. Further, it is readily checked from the expression (\ref{Rexplicit}) of the $\check{R}$-matrix or from examination of the transfer matrix eigenvalues that the spectrum is symmetric under under $S_z\leftrightarrow S_z'$. Therefore, the (hole part of the) spectrum is given by   %
\begin{equation}
\Delta+\bar{\Delta}={\gamma\over 2\pi}\left(S_z^2+(S_z')^2\right)+\ldots \,.
\end{equation}
The central charge is found to be $c=3$ \cite{FJ08} (see also figure~\ref{fig:c_a32RIII} for a numerical check), which  leaves room for only one non-compact degree of freedom. Considerable evidence for this latter degree of freedom has been reported in \cite{VJS2}. 

Based on our earlier result that, under staggering, the perturbation amplitude in the field theory goes as
\begin{equation}
g\propto e^{-{4\over 3}\Lambda} \,,
\end{equation}
we find, using the pole at $\omega={i\pi\over 2}{1\over {\pi\over 4}-\gamma}$, the dimension of the coupling in this regime
\begin{equation}
[g]=[\hbox{length}]^{-{2\over 3}+{8\gamma\over 3\pi}} \,.
\end{equation}
This is compatible with a perturbation of the type
\begin{equation}
g\left\{\partial\varphi\bar{\partial}\varphi \left[e^{i\beta(\phi_1-\phi_2)}+e^{-i\beta(\phi_1+\phi_2)}\right]+e^{2i\beta\phi_2}\right\}
\end{equation}
with the by now familiar correspondence
\begin{equation}
{\beta^2\over 8\pi}={\gamma\over2\pi} \,.
\end{equation}

\subsection{The case $a_4^{(2)}$}
\label{sec:a42RIII}

The ground state is made of 2-strings for both $\lambda^1$-roots and $\lambda^2$-roots, with imaginary parts close to $y_1 = {\pi \over 4} - {3\gamma \over 4}$ and $y_2 = {\pi \over 4} - {\gamma \over 4}$ respectively.
 
After a few manipulations of the Bethe equations, we find the equations for the density of complexes:
\begin{equation}
\rho + \rho^h = \frac{\sinh \omega \left(\frac{\pi}{4} + \frac{5\gamma}{4}\right)-\sinh \omega \left(\frac{\pi}{4} + \frac{\gamma}{4}\right)}{\sinh \frac{\omega\pi}{2}}
- 
\rho 
\frac{\sinh \frac{5}{2}\omega \gamma - \sinh \frac{3}{2}\omega \gamma-\sinh \omega \left(\frac{\pi}{2} -\gamma\right)}{\sinh \frac{\omega\pi}{2}} \,,
\end{equation}
and the solution for the ground state density
\begin{equation}
\rho = \frac{1}{2 \cosh\left(\frac{\omega}{4}(\pi-5 \gamma)\right)} \,.
\end{equation}
In both cases the matrix $K$ is simply a scalar, $K={4\gamma\over\pi}$.

\begin{figure}
\begin{center}
 \includegraphics[scale=0.8]{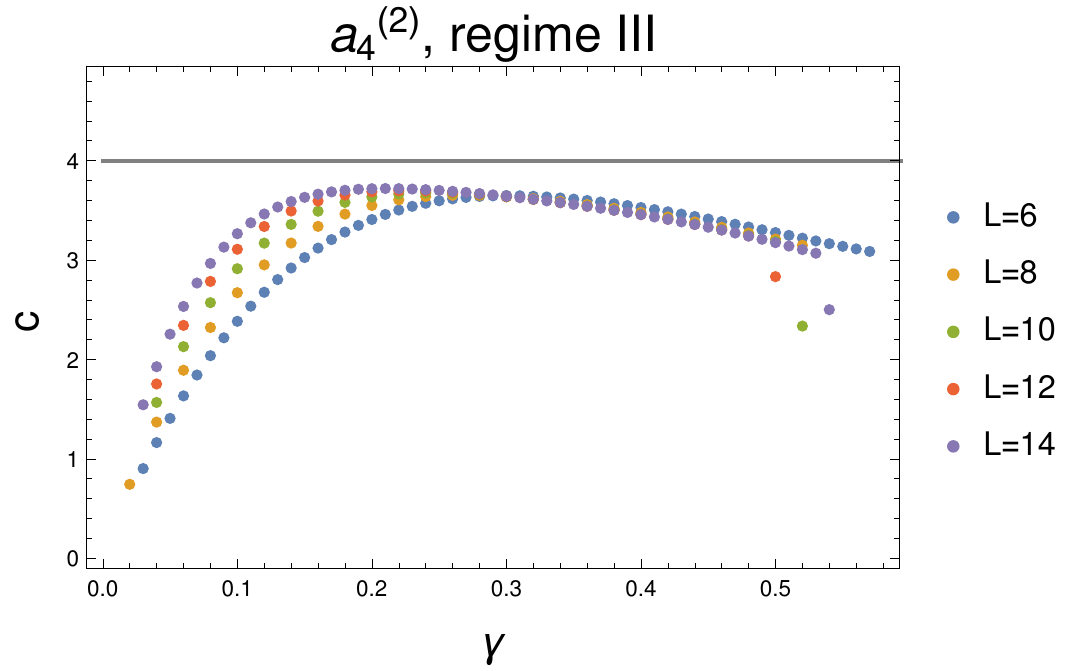}
 \end{center}
\caption{Measure of the central charge in regime III of the $a_4^{(2)}$ model, from numerical solution of the Bethe equations. Important finite-size corrections to the value $c=4$ are observed, associated with the non-compact continuum limit in this regime.}                               
\label{fig:c_a42RIII}                          
\end{figure}

The ground state energy can be written as  
\begin{equation}
 E = \int_{-\infty}^{\infty}\mathrm{d}u \, \frac{-4 \sin \gamma \ \mbox{sech}\left(\frac{2 \pi  u}{\pi -5 \gamma}\right) \left(\cos \gamma-\sin \left(\frac{3\gamma}{2}\right) \cosh (2
   u)\right)}{(\pi -5 \gamma) \left(-2 \left(\sin \left(\frac{\gamma}{2}\right)+\sin \left(\frac{5 \gamma}{2}\right)\right) \cosh (2 u)+\cos
   (2 \gamma)-\cos (3\gamma)+\cosh (4 u)+1\right)}
\end{equation}
with Fermi velocity
\begin{equation}
 v_{\rm F}(\gamma) = \frac{\pi}{\pi - 5 \gamma} \,.
\end{equation}
The central charge can be measured from there; see figure \ref{fig:c_a42RIII}. 
We find $c=4$ with corrections that have the same profile as what we observed in regimes III for $a_2^{(2)}$ and $a_3^{(2)}$, which was characteristic for non-compact degrees of freedom. 

The spectrum associated with holes of complexes is very simple and given by 
\begin{equation}
\Delta+\bar{\Delta}={\gamma\over\pi}n_h^2 \,.
\end{equation}
The existence of two seas of roots suggests the presence of two types of excitations associated with two compact bosons. Thus, the value of the central charge is compatible with the presence of {\em two} extra non-compact degrees of freedom. Unfortunately, we were not able---neither analytically, nor numerically---to obtain reliable information on the corresponding spectra of critical exponents. This will require more study.

\subsection{The case  $a_5^{(2)}$}
\label{sec:a52RIII}


The roots in regime III are found to be arranged as strings over strings over roots, namely 
\begin{eqnarray}
\lambda^1 &\simeq& x \pm \mathrm{i} \left(\frac{\pi}{4}-\gamma\right) \,, \nonumber \\
\lambda^2 &\simeq& x + \mathrm{i} \left(\frac{\pi}{4}-\frac{\gamma}{2}\right) \,, \nonumber \\
\lambda^3 &=& x + \mathrm{i} \frac{\pi}{4} \,.
\end{eqnarray}


There is therefore only one density $\rho$, and the Bethe equations read 
\begin{equation}
\rho + \rho^h = \frac{\sinh \omega \left(\frac{\pi}{4} + \frac{3\gamma}{2}\right)-\sinh \omega \left(\frac{\pi}{4} + \frac{\gamma}{2}\right)}{\sinh \frac{\omega\pi}{2}}
- 
\rho 
\frac{\sinh 3\omega \gamma - \sinh 2\omega \gamma-\sinh \omega \left(\frac{\pi}{2} -\gamma\right)}{\sinh \frac{\omega\pi}{2}} \,,
\end{equation}
with solution for the ground state 
\begin{equation}
\rho = \frac{1}{2  \cosh \frac{\omega}{4}(\pi-6 \gamma)} \,.
\end{equation}

\begin{figure}
\begin{center}
 \includegraphics[scale=0.8]{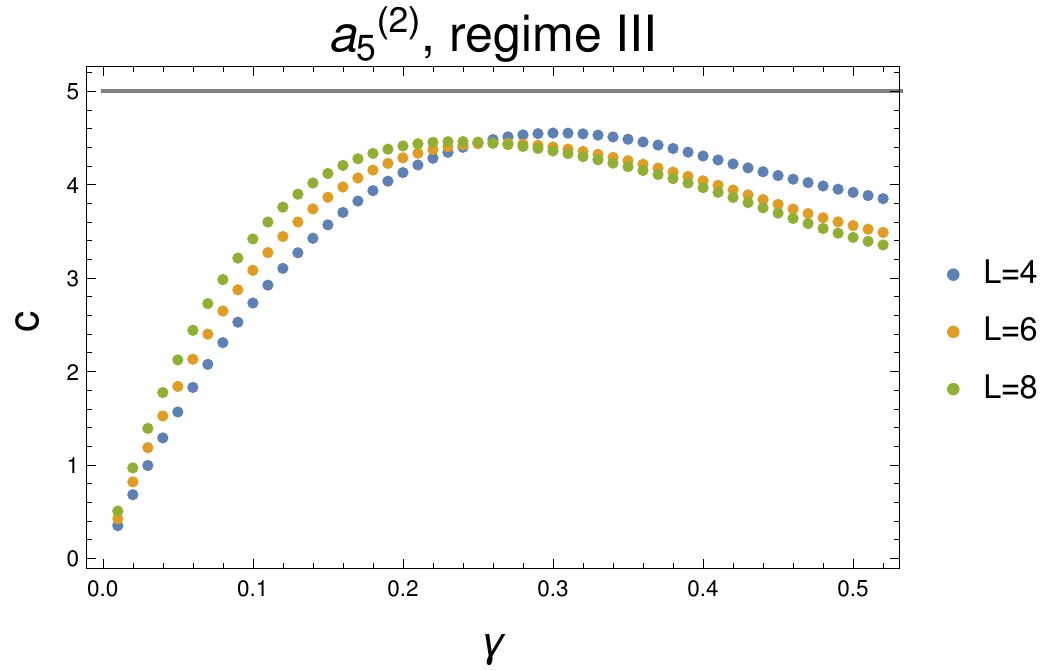}
 \end{center}
\caption{Measure of the central charge in regime III of the $a_5^{(2)}$ model, from numerical diagonalisation of the Hamiltonian (as the numerical resolution of the BAE is in this case very difficult). Important finite-size corrections to the value $c=5$ are observed, associated with the non-compact continuum limit in this regime.
}                               
\label{fig:c_a52RIII}                          
\end{figure}

A numerical estimation of the central charge from data at small system sizes is shown in figure \ref{fig:c_a52RIII}, leading to the conjecture $c=5$, up to large finite-size corrections. 

\subsection{Compact and non-compact sectors}

Regime III corresponds to ${\cal N}>0$ and $\gamma\in [0,{\pi\over 2n}]$ for $a_{2n-1}^{(2)}$, resp.\ $\gamma\in [0,{\pi\over 2n+1}]$ for $a_{2n}^{(2)}$. Setting $\gamma\equiv {\pi\over N+k-2}$ for $a_{N-1}^{(2)}$, we know that for $k$ integer, the quantum group restricted model is the conformal coset $SO(N)_k/SO(N-1)_k$. It is natural to expect that the continuum limit of the untwisted model be related with the Coulomb gas description of these cosets, elements of which are discussed in the appendix. Some features follow immediately;  in particular, we obtain further elements that confirm the pattern previously established:
$$
\begin{array}{l|l|l|l}
a_{2n-1}^{(2)} & \hbox{Rank $n$} &
\left \lbrace \begin{array}{ll} \mbox{$n$ compact bosons} \\ \mbox{$n-1$ non-compact bosons} \end{array} \right.
& c=2n-1\\ \hline
a_{2n}^{(2)} & \hbox{Rank $n$} & \left\lbrace \begin{array}{ll} \mbox{$n$ compact bosons} \\ \mbox{$n$ non-compact bosons} \end{array} \right.
& c=2n
\end{array}
$$
The number of compact bosons is the same as what we have found in regime I. However, it is important to observe that the compactification lattices are different. In the case of regime I, the vertex operators $e^{i{\pmb \alpha} \cdot {\pmb \phi}}$ had `charges' ${\pmb \alpha}$ belonging to the root lattice of 
$b_n$ for $a_{2n}^{(2)}$ and $c_n$ for $a_{2n-1}^{(2)}$. In regime III, these charges should  belong instead to the weight lattice, if the coset interpretation is correct.  Moreover, for $a_{2n-1}^{(2)}$, we expect the emergence of the $d_n$ (and not $c_n$) weight lattice, a feature to be investigated further.%
\footnote{Recall that while the root lattices of $c_n$ and $d_n$ are the same, the weight lattices are different. Up to a normalisation, one can always write the weight lattice as $\pmb{ \Lambda}(c_n)=Z[{\pmb e}_1,\ldots,{\pmb e}_n]$ and $\pmb{\Lambda}(d_n)=Z[{\pmb e}_1,\ldots,{\pmb e}_n;({\pmb e}_1+\ldots+{\pmb e}_n)/2]$, where $Z[\cdots]$ denotes the span over the integers. For $n=2$ for instance, if  the weight lattice of $c_2$ is a square lattice, the weight lattice of $d_2$ is another square lattice, rotated by a $45$ degree angle and contracted by a factor $\sqrt{2}$. While our results for $a_3^{(2)}$ are compatible with this, we were unfortunately unable to explore the question for higher values of $n$. }

Like in the cases of low rank that we discussed in details before, the $a_{N-1}^{(2)}$ chains in regime III again provide, after staggering, an integrable lattice discretisation of  integrable massive  perturbed CFTs. It is easy to speculate what this theory might be. Indeed, 
we saw that the continuum limit of the staggered $a_{N-1}^{(2)}$ model in regime III can be described (for appropriate values of $\gamma$, and after quantum group reduction) as an $SO(N)/SO(N-1)$ gauged WZW model, with the numerator and denominator at level $\widetilde{N}$ when
\begin{equation}
\gamma={\pi\over N+\widetilde{N}-2} \,.
\end{equation}
There are two well-known integrable perturbations of such gauged WZW models \cite{Miramontes} . The one we are interested in involves a perturbation with weight  $h={N\over N+\widetilde{N}-2}$. In general, the conformal weights in the $SO(N)_{\tilde{N}}$ WZW models are 
\begin{equation}
h={C_2\over N+\tilde{N}-2} \,,
\end{equation}
thus the perturbation corresponds to a representation with the Casimir $C_2$  equal to $N$ for $SO(N)$. If the weight reads $\pmb{\lambda}=\sum_{i=1}^n \lambda_i \pmb{e}_i$ (the ${\pmb e}_i$ are as usual a set of orthonormal vectors), the Casimirs read
\begin{eqnarray}
C_2 &=& \pmb{\lambda} \cdot (\pmb{\lambda}+2\pmb{\rho})=\sum_{i=1}^n \left[\lambda_i^2+(2n+1-2i)\lambda_i\right] \,, \quad \ \mbox{for }SO(2n+1) \,, \nonumber\\
C_2 &=& \pmb{\lambda} \cdot (\pmb{\lambda}+2\pmb{\rho})=\sum_{i=1}^n \left[\lambda_i^2+(2n-2i)\lambda_i\right] \,, \qquad \quad \mbox{for }SO(2n) \,, \label{casimirs}
\end{eqnarray}
so we see that the choice $\lambda_1=2$ and $\lambda_i=0$ otherwise gives  $C_2=N$ for $SO(N)$. The conformal weight in the untwisted theory is then
\begin{equation}
\Delta+\bar{\Delta}= {\pmb{\lambda} \cdot \pmb{\lambda}\over k+g}={4\over N+\widetilde{N}-2}={4\gamma\over\pi} \,,
\end{equation}
where $k$ has the same meaning as in section~\ref{sec:duality} and $g$ is the dual Coxeter number.
This value of $\Delta + \bar{\Delta}$ corresponds to a number of holes $n_h=2$ in the finite-size spectrum.

We can finally write down the perturbed theory  in terms of the free fields identified at the critical point. We have seen earlier that 
\begin{equation}
g\propto e^{-{4\over H}\Lambda} \,,
\end{equation}
a relation that is independent of the regime. This, together with the analysis of the Bethe equations for the staggered chain, suggest the following perturbation in the case $a_{2n-1}^{(2)}$:
\begin{equation}
g\left\{\partial\varphi_1\bar{\partial}\varphi_1\left[e^{i\beta(\phi_{1}-\phi_2)}+e^{-i\beta(\phi_{1}
+\phi_2)}\right]+2\sum_{j=2}^{n-1}\partial\varphi_j\bar{\partial}\varphi_j
e^{i\beta(\phi_j-\phi_{j+1})}+e^{2i\beta\phi_n}\right\}\label{Todageniiagain}
\end{equation}
with $n$ compact bosons $\phi_j$ and $n-1$ non-compact ones $\varphi_j$. The action (\ref{Todageniiagain}) is of course   an $a_{2n-1}^{(2)}$ Toda theory coupled to  non-compact bosons. 
Similar results are obtained for $a_{2n}^{(2)}$.

\section{Conclusion}
\label{sec:conc}

To summarise, we have found that the low-energy limit of the $a_2^{(2)}$ spin chain can be described, depending on the regime, by the UV limit of three different integrable massive QFT:
\begin{enumerate}
\item In regime I, with ${\cal N}<0$ and $\gamma\in[0,\pi]$:
\begin{equation}
 S = S_{\rm FBc} + g\int {\rm d}^2z \, \left[e^{-2i\beta\phi}+e^{i\beta\phi}\right] \,;
\end{equation}
\item In regime II, with ${\cal N}>0$ and $\gamma\in[{\pi\over 3},\pi]$:
\begin{equation}
S = S_{\rm FBc} + S_{\rm Ising}+g\int {\rm d}^2z \, \left[e^{-2i\beta\phi}+\psi\bar{\psi}e^{i\beta\phi}\right] \,;
\end{equation}
\item In regime III, with ${\cal N}>0$ and $\gamma\in[0,{\pi\over 3}]$:
\begin{equation}
 S=S_{\rm FBc}+S_{\rm FBnc}+g\int {\rm d}^2z \, \left[e^{-2i\beta\phi}+\partial\varphi\bar{\partial}\varphi e^{i\beta\phi}\right] \,,
 \end{equation}
\end{enumerate}
where, in obvious notations, $S_{\rm FBc}$ (resp.\ $S_{\rm FBnc}$ ) denotes the free boson action for a compact boson $\phi$ (resp.\ a non-compact boson $\varphi$), and $S_{\rm Ising}$ is the action for a free Majorana fermion $\psi$. In the third equation, note the coupling between the compact boson $\phi$ and the non-compact boson $\varphi$. This pattern essentially generalises to the case of $a_{2n}^{(2)}$ with $n > 1$. For $a_{2n-1}^{(2)}$ the intermediate regime II with fermions is not observed. The general result for $a_{N-1}^{(2)}$ in all three regimes---including the extent of the regimes and the number of compact and non-compact degrees of freedom---is summarised in table~\ref{summ}.

The emergence of a series of non-compact CFTs is of course fascinating, and requires much more work to be thoroughly understood. In particular, all we have done is to give evidence for the counting of compact and non-compact degrees of freedom. This is far from a whole description of the CFTs. Like in the case of the $a_2^{(2)}$ model \cite{VJS1} one would like to have an understanding of these theories in terms of a sigma model (like the Euclidean black hole theory \cite{Witten91,Dijkgraaf92}) or some generalisation of the (dual, for $a_2^{(2)}$) sine-Liouville theory. One would like also to know the density of states for the continuous part of the spectrum (this was partially achieved \cite{IJS12} for $a_2^{(2)}$, but via a different lattice regularisation \cite{JS06,IJS08}) and whether the coset models can be obtained by a projection onto the set of discrete states (like for $a_2^{(2)}$). Finally, one would like to understand the  properties of the integrable massive deformations. It should take quite a while to complete this program.
 
In any case, the systematic emergence of a continuum limit with non-compact degrees of freedom raises the general question of what might happen in less explored regimes of other spin chains. It is intriguing to note in this respect that the staggering of the $a_{N-1}^{(2)}$ spin chains produces $SO(N)/SO(N-1)$ perturbed gauged WZW models, while, on the other hand, these models are well-known to be related with the  {\sl Pohlmeyer reduction} \cite{Miramontes} of 
 $SU(N)/SO(N)$ sigma models.%
 \footnote{Recall that in general the Pohlmeyer reduction involves a triplet of Lie groups, $H\subset G\subset F$, a sigma model on $F/G$, and a perturbed CFT on $G/H$.}
 The presence of $SU(N)$ in the numerator is of course related with the underlying $a_{N-1}$ structure of twisted $a_{N-1}^{(2)}$ theories. Now there are many more integrable perturbed gauged WZW models, and the natural  question is whether they can also be obtained as the continuum limit of some spin chains. For instance, the reduction of $SO(N+1)/SO(N)$ is also associated with the $SO(N)/SO(N-1)$ gauged WZW models but perturbed by a different field. For $N=3$, the case we have studied here corresponds  to  $SU(3)/SO(3)$ and parafermions perturbed by the second energy operator, while $SO(4)/SO(3)$ would correspond instead to parafermions perturbed  by the first energy operator. Now a  lattice regularisation is in fact known for the latter case \cite{Ikhlef}, suggesting at the very least the existence of another family of lattice models whose continuum limit would be the same theories we have found here, but whose staggering would lead to a different perturbation (presumably with $\lambda_1=1$ in (\ref{casimirs})). This will be discussed elsewhere.
 
 Yet another interesting question concerns the emergence of different cosets in the continuum limits of spin chains. While initial studies on the $SU(N)$ case produced only cosets $SU(N)\times SU(N)/SU(N)$, it is natural to wonder now whether there exists other chains producing other cosets, for instance $SU(N)/SU(N-1)$. The question can be asked both for the RSOS versions, and for the corresponding `Coulomb gas' interpolations. This question, too, will be discussed elsewhere. 
 
\bigskip
\noindent {\bf Acknowledgments}: We thank V.\ Fateev, A.\ Kuniba and R.\ Nepomechie for useful discussions. This work was supported by the ANR DIME, the ERC Advanced Grant NuQFT, and the Institut Universitaire de France. EV acknowledges support by the ERC Starting Grant 279391 EDEQS.

\appendix

\section{Free-field representation of the cosets $SO(N)_k/SO(N-1)_k$}
\label{sec:app}

Following for instance the paper \cite{kuwahara}, we can bosonise the $SO(2n)$ model with $n(n-1)$ pairs $\beta_i, \gamma_i$ and $n$ free scalar fields. Meanwhile, we bosonise the $SO(2n+1)$ model 
with $n^2$ pairs $\beta'_j,\gamma'_j$ and $n$ scalar fields. 

If we take the coset $SO(2n+1)/SO(2n)$, we thus get $n$ pairs $\beta,\gamma$---that is, $n$ compact bosons and $n$ non-compact ones. Note that the dimension $D$ of $SO(N)$ is $N(N-1)/2$, so $D[SO(2n+1)]-D[SO(2n)]=2n$.

If we take the coset $SO(2n+2)/SO(2n+1)$, we get $n$ pairs $\beta,\gamma$ and $1$ scalar field---that is, $n+1$ compact bosons and $n$ non-compact ones.  Note that $D[SO(2n+2)]-D[SO(2n+1)]=2n+1$. So this fits.

Let us now give a few more details. The group  $SO(2n+1)$ has dimension $n(2n+1)$ and dual Coxeter number $g=2n-1$. Introducing the usual orthonormal basis $\pmb{e}_i$, with $i=1,2,\ldots,n$, we have the roots $\pmb{\alpha}_{ij}\equiv \pmb{e}_i-\pmb{e}_j$,  ($i\neq j$), the roots $\pm \pmb{\delta}_{ij}$ with $\pmb{\delta}_{ij}=\pmb{e}_i+\pmb{e}_j$ ($i\neq j$), and the roots $\pm \pmb{e}_i$. For the currents, we use a Wakimoto construction, which requires for the first type of roots ${n(n-1)\over 2}$ pairs of $\beta_{ij},\gamma_{ij}$ bosons (we take by convention $i>j$ for positive roots), ${n(n-1)\over 2}$ pairs of $\widetilde{\beta}_{ij},\widetilde{\gamma}_{ij}$ bosons for the second type of roots, and $n$ pairs of $\beta_i,\gamma_i$ for the third type of roots. Finally, we introduce $n$ bosons $\phi_i$ for the Cartan generators. The corresponding stress energy tensor reads
\begin{equation}
T_{SO(2n+1)}=\sum_{i>j}\beta_{ij}\partial\gamma_{ij}+\widetilde{\beta}_{ij}\partial\widetilde{\gamma}_{ij}+
\sum_i\beta_i\partial\gamma_i-{1\over 2}\sum_i (\partial\phi_i)^2-{i\over\alpha_G}\pmb{\rho}_G \cdot \partial^2\pmb{\phi} \,,
\end{equation}
where $\alpha_+\equiv\sqrt{k+2n-1}$ and the last term involves  the usual half-sum of positive roots
\begin{equation}
\pmb{\rho}_G\equiv \sum_{i=1}^n \left(n-i+{1\over 2}\right)\pmb{e}_i \,.
\end{equation}
For $SO(2n)$ we have dimension $n(2n-1)$ and the dual Coxeter number $g=2n-2$. The set of roots is the same, except for the last type $\pm \bf{e}_i$ which are now absent. The Cartan sub-algebra is generated by fields $\phi'_i$, with the stress tensor
\begin{equation}
T_{SO(2n)}=\sum_{i>j}\beta_{ij}\partial\gamma_{ij}+\tilde{\beta}_{ij}\partial\tilde{\gamma}_{ij}-{1\over 2}\sum_i (\partial\phi'_i)^2-{i\over\alpha_H}\pmb{\rho}_H \cdot \partial^2\pmb{\phi}'
\end{equation}
with now
\begin{equation}
\pmb{\rho}_H=\sum_{i=1}^n (n-i)\pmb{e}_i \,.
\end{equation}
The coset construction involves a sum over  the positive roots in $G/H$, which in this case gives simply
\begin{eqnarray}
T_{G/H} &=&\sum_j \beta_j\partial\gamma_j-{1\over 2}\sum_j (\partial\phi_j)^2-{i\over\sqrt{k+2n-1}}\sum_j \left(j-{1\over 2}\right)\partial^2\phi_j\nonumber\\
& & +{1\over 2}\sum_j (\partial\phi'_j)^2+{i\over\sqrt{k+2n-2}}\sum_j (j-1)\partial^2\phi'_j \,,
\end{eqnarray}
while the identification of the Cartan generators imposes the constraints
\begin{equation}
\sqrt{k+2n-1}\partial\phi_j'=\sqrt{k+2n-2}\partial\phi_j+i\beta_j\gamma_j \,.
\end{equation}
In all these expressions, the label $j$ runs over $j=1,2,\ldots,n$. We now bosonise the $\beta,\gamma$ systems by setting
\begin{equation}
\gamma_j\equiv e^{-\Phi+\chi},~\beta_j\equiv \partial\chi_j e^{\Phi-\chi} \,,
\end{equation}
so we have 
\begin{eqnarray}
\beta\gamma &=& \partial\Phi \,, \nonumber\\
\beta\partial\gamma &=& -{1\over 2}(\partial\Phi)^2+{1\over 2}(\partial\chi)^2-{1\over 2}\partial^2\chi+{1\over 2}\partial^2\Phi \,,
\end{eqnarray}
and the relation between the Cartans becomes
\begin{equation}
\sqrt{k+2n-1}\partial\phi_j'=\sqrt{k+2n-2}\partial\phi_j+i\partial\Phi_j \,.
\end{equation}
Introduce now 
\begin{eqnarray}
\sigma_j &\equiv& i\sqrt{k+2n-2}\chi_j+i{\sqrt{k+2n-1}\over\sqrt{k+2n-2}}\Phi_j+\sqrt{{k+2n-1\over k+2n-2}}\phi_j \,, \nonumber\\
\nu_j &\equiv& i\sqrt{k+2n-1}\chi_j+i\sqrt{k+2n-1}\Phi_j+\phi_j \,.
\end{eqnarray}
The stress-energy tensor of the coset theory can then be written:
\begin{eqnarray}
T &=&\sum_{j=1}^n {1\over 2}(\partial\sigma_j)^2-{1\over 2}(\partial\nu_j)^2\nonumber\\
& & + i{j-k-2n+1\over\sqrt{k+2n-2}}\partial^2\sigma_j-i{j-k-2n+1/2\over \sqrt{k+2n-1}}\partial^2\nu_j \,.
\end{eqnarray}
The propagators are
\begin{eqnarray}
\langle \nu_j(z)\nu_k(w)\rangle &=& -\delta_{jk}\ln(z-w) \,; \nonumber \\
\langle \sigma_j(z)\sigma_k(w)\rangle &=& +\delta_{jk}\ln(z-w) \,.
\end{eqnarray}
The first immediate observation is that the untwisted theory is a set of $n$ compact and $n$ non-compact bosons. The conformal weights of the  twisted theory should reproduce, after the introduction of screening charges etc, the conformal weights of the coset CFT 
\begin{equation}
h={\pmb{\lambda} \cdot (\pmb{\lambda}+2\pmb{\rho}_G)\over 2(k+g)}-
{\pmb{\mu} \cdot (\pmb{\mu}+2\pmb{\rho}_H)\over 2(k+h)} \,.
\end{equation}
where $\pmb{\lambda}$ (resp. $\pmb{\mu}$) belongs to the weight lattice of $G$ (here, $SO(2n+1)$) (resp. $H=SO(2n)$). These conformal weights should appear via discrete states in the model. The untwisted model meanwhile will have a spectrum made of a discrete part (the $G$ part) and a continuous one---i.e., it should have the form
\begin{equation}
\Delta={\pmb{\lambda}.\pmb{\lambda}\over 2(k+g)}+\hbox{continuum} \,.
\end{equation}

\end{document}